\documentclass[showpacs,preprint,preprintnumbers,aps,nofootinbib]{revtex4-1}

\usepackage{graphicx}
\usepackage{dcolumn}
\usepackage{bm}
\usepackage{amsmath,amssymb,amsfonts}
\usepackage{latexsym}
\usepackage{ulem}

\usepackage{color}

\def\nn    {\nonumber}
\def\dcp{\Delta \mathcal{A}_{\text{CP}}}
\def\ebsg{$B\to X_s \gamma$~}
\def\bsg{\mathcal{B}(B\to X_s \gamma)}
\def\rbb{\rho_{bb}}
\def\rtt{\rho_{tt}}
\def\irbb{\mbox{Im}(\rho_{bb})}
\def\rrbb{\mbox{Re}(\rho_{bb})}

\def\fbi{fb$^{-1}$~} 
\def\bgba{$bg \to b A \to b t \bar t$~}
\def\bzh{$bg \to b A \to b Z h$~}
\def\gga{$gg \to t \bar t A \to t \bar t b \bar b$~}

\begin{document}

\preprint{OU-HET-1061}
\title{\boldmath
Probing Electroweak Baryogenesis induced by extra bottom Yukawa coupling via EDMs and collider signatures}

\author{Tanmoy Modak$^{1,2}$ and Eibun Senaha$^{3,4}$}
\affiliation{$^1$Department of Physics, Osaka University, Toyonaka, Osaka 560-0043, Japan \\
$^2$Department of Physics, National Taiwan University, Taipei 10617, Taiwan\\
$^3$Theoretical Particle Physics and Cosmology Research Group, Advanced Institute of Materials 
Science, Ton Duc Thang University, Ho Chi Minh City, Vietnam\\
$^4$Faculty of Applied Sciences, Ton Duc Thang University, Ho Chi Minh City, Vietnam}
\bigskip

\begin{abstract} 
We study the prospect of probing electroweak baryogenesis driven by an extra bottom Yukawa coupling
$\rbb$ in a general two Higgs doublet model via electric dipole moment (EDM) measurements and at the 
collider experiments. The parameter space receives meaningful constraints from 
125 GeV Higgs $h$ boson signal strength measurements as well as several heavy Higgs boson searches 
at the Large Hadron Collider (LHC). In addition, we show that the asymmetry of the CP asymmetry of 
inclusive $B\to X_s \gamma$ decay would provide complementary probe. A discovery is possible at the LHC
via $bg\to bA \to b Z h$ process if $|\rbb|\sim 0.15$ and $250~\mbox{GeV}\lesssim m_A \lesssim 350$ GeV,
where $A$ is CP odd scalar. For $m_A> 2 m_t$ threshold, where $m_t$ is the top quark mass,
one may also discover $bg\to bA \to b t \bar t$ at the high luminosity LHC run if an extra 
top Yukawa coupling $|\rtt|\sim0.5$, though it may suffer from systematic uncertainties. For completeness we study 
$gg\to t \bar t A \to t \bar t b \bar b $ but find it not promising.
\end{abstract}

\maketitle

\section{Introduction}
The existence of the matter-antimatter asymmetry is unswervingly established over the years by various 
cosmological observations such as the cosmic microwave 
background anisotropies and big-bang nucleosynthesis~\cite{Tanabashi:2018oca}.
It has been understood that the Universe started with 
equal number of baryons and antibaryons, but later evolved into baryon 
dominated Universe dynamically via a mechanism called baryogenesis. 
A successful baryogenesis requires three necessary conditions namely, baryon number violation, 
charge conjugation (C) and charge conjugation-parity (CP) violation and, 
departure from thermal equilibrium, laid out by Sakharov in 1967~\cite{Sakharov:1967dj}.
A plethora of baryogenesis scenarios have been proposed so
far to account for the observed baryon asymmetry of the Universe (BAU), however, 
its origin is still unclear. After the discovery of the 125 GeV Higgs boson at the Large Hadron 
Collider (LHC)~\cite{h125_discovery}, a significant 
attention has been directed in particular to electroweak baryogenesis
(EWBG)~\cite{ewbg,ewbg_2hdm,ewbg_susy,ewbg_singletSM,ewbg_others,ewbg_g2HDM,Fuyuto:2017ewj,Modak:2018csw,Fuyuto:2019svr,highewbg} 
mechanism for its direct connections to Higgs physics and, its testability at the ongoing experiments. 
The Standard Model (SM) belongs to this class, however the CP violation
is too small and, the electroweak symmetry breaking is not strongly
first order phase transition (EWPT) to drive departure from thermal equilibrium.

While we do not have any strong experimental evidence of new physics yet, multi-Higgs sector
is the natural consequence of most ultraviolet (UV) theories due to enlarged symmetries. 
Whatever the fundamental theory might be, their effective descriptions at $\mathcal{O}(100)$ 
GeV scale should resemble the SM in light of the latest experimental results. As for the Higgs
sector, two cases are conceivable: one is that all the new scalars are much heavier than
$\mathcal{O}(100)$ GeV scale, thereby the Higgs sector is effectively reduced to the SM, 
while the other is that new scalars have $\mathcal{O}(100)$ GeV masses but their couplings
to the gauge bosons and fermions are SM like, mimicking the SM. From the viewpoint of new 
physics discovery potential, it is timely to consider the latter case and investigate 
whether the aforementioned cosmological issue can be solved or not. 
Since we have already confirmed the existence of the Higgs doublet in nature, 
it is tempting to us to think of additional Higgs doublets 
in analogy with the fact that all the fermions come in three copies.

The general two Higgs doublet model (g2HDM) is one of the simplest renormalizable low-energy models where the 
scalar sector of the SM is extended by an extra scalar doublet~\cite{2hdmreview}.
Without the presence of discrete symmetry, in g2HDM, both the scalar 
doublets couple with up- and down-type fermions at tree level. In the mass eigenbasis of the fermions ($F$),
one has two independent Yukawa couplings $\lambda^F_{ij}$ and $\rho^F_{ij}$, where
the former is real and diagonal that are responsible for the fermion mass generation, while the latter is complex and non-diagonal.
Such complex couplings can  provide additional CP violating sources 
beyond the usual Cabibbo-Kobayashi-Maskawa (CKM)  framework~\cite{ckm} of the SM.

EWBG in g2HDM is widely investigated in Refs.~\cite{ewbg_g2HDM,Fuyuto:2017ewj,Modak:2018csw,Fuyuto:2019svr}. This model can simultaneously accommodate 
the strong first-order EWPT and sufficient amount of CP violation which the SM fails to provide. 
The most natural EWBG scenario in g2HDM would be the case in which BAU is driven by the extra top
Yukawa coupling ($\rho_{tt}$) of $\mathcal{O}(0.01-1)$ in magnitude with moderate size of the CP 
phase ($\rho_{tt}$-EWBG)~\cite{Fuyuto:2017ewj}. The devoted collider study of this scenario is 
conducted in Ref.~\cite{Kohda:2017fkn}. 

As a complementary study to $\rho_{tt}$-EWBG, the present authors consider a scenario 
in which the CP phase of $\rho_{tt}$ is approximately zero and the extra bottom 
Yukawa coupling ($\rho_{bb}$) plays a dominant role in generating BAU ($\rho_{bb}$-EWBG)~\cite{Modak:2018csw}. 
It is demonstrated that BAU can reach the observed level if $|\text{Im}\rho_{bb}|\gtrsim0.058$
with generous assumptions on a Higgs bubble wall profile. 
There exist several direct and indirect search constraints on the parameter space for $\rho_{bb}$-EWBG
such as $h$ boson signal strength measurements, heavy Higgs searches at the LHC. 
The $\rho_{bb}$-EWBG can be discovered at the LHC via $bg\to b A \to b ZH$ (or $bg\to b H \to b ZA$) process if 
$|\irbb| \sim \mathcal{O}(0.1)$~\cite{Modak:2019nzl}. However, the process requires that
$m_A > m_H + m_Z$ and $\rtt$ to be negligibly small to avoid constraints from flavor physics~\cite{Modak:2019nzl}. In addition, 
for $m_A > 2 m_t$, the process gets dilution from $A \to t \bar t$ decay if $\rtt$ is nonvanishing.
Also, it would be extremely difficult to probe the phase of $\rho_{bb}$ at the LHC since its information is lost in $pp$ collision.

In this paper we show that the electron EDM measurement and asymmetry 
of CP asymmetry of the $B\to X_s \gamma$ decay offer exquisite probes for $\irbb$. 
We also analyze the prospect of discovery at the LHC. In particular we study
the discovery potential of $\rho_{bb}$-EWBG via \bzh and \bgba
processes at 14 TeV LHC. Purpose of this paper is to find possible direct and indirect signatures 
and correlation between them in probing the parameter space for the $\rho_{bb}$-EWBG.

Induced by $\rbb$, the \bzh process can be searched at the LHC via $pp \to b A + X \to b Z h +X$
($X$ is inclusive activities) followed by $Z\to \ell^+ \ell^-$ ($\ell=e,~\mu$) and $h\to b\bar b$ decays,
constituting same flavor opposite sign dilepton pair and three $b$-tagged jets.
While the \bzh process can be induced $\rbb$, the \bgba process requires both $\rho_{bb}$ and $\rtt$ to be nonvanishing.
The latter process can be searched via $pp \to b A + X \to b t \bar t +X$ with at least one top decays semileptonically,
constituting three $b$-tagged jets, at least one charged lepton ($e$ and $\mu$) and missing transverse energy signature 
(denoted as $3b1\ell$ process). These processes provide the sensitive probes 
for the parameter space of $\rho_{bb}$-EWBG, which is complementary to Ref.~\cite{Modak:2019nzl}.

For the sake of completeness we also investigate the discovery prospect of the
$gg \to t \bar t A \to t \bar t b \bar b$ process, which is induced by nonzero $\rho_{bb}$ and $\rtt$. 
At the LHC the process can be searched via $pp\to t \bar t A + X \to t \bar t b \bar b + X$,
with at least one top decaying semileptonically. As $\rho_{tt}$ gets involved in both 
\bgba and $gg \to t \bar t A \to t \bar t b \bar b$, the processes would provide complementary 
probes also for $\rho_{tt}$-EWBG. 

In the following, we outline the formalism in Sec.~\ref{frame},
followed by a detailed discussion on the available parameter space and potential indirect probes in Sec.~\ref{param}. 
We discuss the discovery potential of $\rho_{bb}$-EWBG at the LHC in Sec.~\ref{coll} and
summarize our results with some outlook in Sec.~\ref{summ}.

\section{Framework}\label{frame}
The particle content of g2HDM is the SM plus additional Higgs doublet. 
In general, this model induces flavor-changing neutral current (FCNC) processes mediated
by the neutral Higgs bosons at tree level. It is common to impose a $Z_2$ symmetry to suppress
the FCNC processes to be consistent with various flavor physics data. Though this setup works well, 
having the $Z_2$ symmetry implies that the model has some specific UV theories such as supersymmetric 
models. Since we do not try to connect the model to any specific UV completions, we do not impose 
the $Z_2$ symmetry or something similar, which enables us to discuss physics at $\mathcal{O}(100)$ 
GeV scale in wider perspective. In this bottom-up approach, the tree-level FCNC processes are possible 
as long as the experimental data allow, and sources of CP violation are much richer than 2HDMs with some discrete symmetries. 

The most general two Higgs doublet potential can be 
written in the Higgs basis as~\cite{Davidson:2005cw, Hou:2017hiw}
\begin{align}
V(\Phi,\Phi') &= \mu_{11}^2|\Phi|^2 + \mu_{22}^2|\Phi'|^2 - (\mu_{12}^2\Phi^\dagger\Phi' + \text{H.c.})
 + \frac{\eta_1}{2}|\Phi|^4 + \frac{\eta_2}{2}|\Phi'|^4 + \eta_3|\Phi|^2|\Phi'|^2\nn\\
 &\quad +\eta_4 |\Phi^\dagger\Phi'|^2 + \bigg[\frac{\eta_5}{2}(\Phi^\dagger\Phi')^2
     + \left(\eta_6 |\Phi|^2 + \eta_7|\Phi'|^2\right) \Phi^\dagger\Phi' + \text{H.c.}\bigg].
\label{pot}
\end{align}
Each Higgs doublet fields is expressed as
\begin{align}
\Phi = 
\begin{pmatrix}
G^+ \\
\frac{1}{\sqrt{2}}(v+h+iG^0)
\end{pmatrix},\quad
\Phi' = 
\begin{pmatrix}
H^+ \\
\frac{1}{\sqrt{2}}(A+H)
\end{pmatrix},
\end{align}
where $v(\simeq 246~\text{GeV})$ is the vacuum expectation value, $h$ is the SM-like Higgs boson, $G^{0,\pm}$ are the Nambu-Goldstone bosons,
$H$ and $A$ are the CP-even and -odd Higgs bosons, respectively, and $H^\pm$ are the charged Higgs bosons. From the minimization condition with 
respect to $\Phi$, it follows that $\mu_{11}^2=-\eta_1 v^2/2$. For simplicity, we assume CP-conserving Higgs sector at tree level.\footnote{Since 
we have CP violation in the Yukawa sector as delineated below, its effect appears in the Higgs spectrum at one-loop level and CP-even and 
-odd Higgs boson mix with each other. Nevertheless, such a one-loop induced mixing
is so small that $\{h,H,A\}$ can be regarded as the mass eigenstates to a good approximation.} 
The second minimization condition with respect to $\Phi'$ gives $\mu_{12}^2 = \eta_6 v^2/2$.

The mixing angle $\gamma$ between the CP-even bosons $h$ and $H$
satisfies the relations~\cite{Hou:2017hiw}
\begin{align}
 \cos^2\gamma = \frac{\eta_1 v^2 - m_h^2}{m_H^2-m_h^2},~\quad \quad \sin{2\gamma} = \frac{2\eta_6 v^2}{m_H^2-m_h^2}.
\end{align} 
An alignment limit is defined as $c_\gamma = 0$ and $s_\gamma = -1$, where $c_\gamma$ and $s_\gamma$ are shorthands for 
$\cos\gamma$ and $\sin\gamma$ respectively.
One can express the masses of $h$, $H$, $A$ and $H^\pm$ in terms of the parameters in Eq.~\eqref{pot}:
\begin{align}
 m_{h,H}^2 &= \frac{1}{2}\bigg[m_A^2 + (\eta_1 + \eta_5) v^2\mp \sqrt{\big[m_A^2+ (\eta_5 - \eta_1) v^2\big]^2 + 4 \eta_6^2 v^4}\bigg],\\
 m_{A}^2 &= \frac{1}{2}(\eta_3 + \eta_4 - \eta_5) v^2+ \mu_{22}^2,\\
 m_{H^\pm}^2 &= \frac{1}{2}\eta_3 v^2+ \mu_{22}^2.
\end{align}
Note that in the alignment limit, one has $m_h^2 = \eta_1v^2$ and 
$m_H^2 = m_A^2+\eta_5v^2=(\eta_3+\eta_4+\eta_5)v^2/2+\mu_{22}^2$. In contrast to $m_h$, the masses 
of the extra Higgs bosons are controlled by $\eta_i v^2$ and $\mu_{22}^2$, where $\eta_i$ denotes some linear 
combinations of the $\eta$ couplings. As is well known, magnitudes of the heavy Higgs loop contributions can 
become sizable if $\eta_iv^2\gtrsim \mu_{22}^2$, which is necessary for achieving the strong first-order EWPT.

The CP-even scalars $h$, $H$ and CP-odd scalar $A$ couple to fermions 
by~\cite{Davidson:2005cw}
\begin{align}
\mathcal{L} &= 
-\frac{1}{\sqrt{2}} \sum_{F = U, D, L}
 \bar F_{i} \bigg[\big(-\lambda^F_{ij} s_\gamma + \rho^F_{ij} c_\gamma\big) h 
 +\big(\lambda^F_{ij} c_\gamma + \rho^F_{ij} s_\gamma\big)H -i ~{\rm sgn}(Q_F) \rho^F_{ij} A\bigg]P_R\; F_{j}\nn\\
 &\quad-\bar{U}_i\left[(V\rho^D)_{ij} P_R-(\rho^{U\dagger}V)_{ij} P_L\right]D_j H^- \bar{\nu}_i\rho^L_{ij} P_R \; L_j H^+ +{\rm H.c.},\label{eff}
\end{align}
where $P_{L,R}\equiv (1\mp\gamma_5)/2$, $i,j = 1, 2, 3$ are generation indices, $V$ is CKM matrix,
and $U=(u,c,t)$, $D = (d,s,b)$, $L=(e,\mu,\tau)$ and $\nu=(\nu_e,\nu_\mu,\nu_\tau)$ are in vectors in flavor space. 
The matrices $\lambda^F_{ij}\; (=\sqrt{2}m_i^F/v)$ are real and diagonal,
whereas $\rho^F_{ij}$ are in general complex and non-diagonal.

Purpose of this paper is to probe the parameter space for
EWBG driven by the extra bottom Yukawa $\rbb$. It is found that a successful EWBG requires $|\irbb|\gtrsim 0.058$~\cite{Modak:2018csw}.
In the following we shall show that the parameter space receives
meaningful constraints from several direct and indirect searches. The most stringent constraint on $\irbb$ arises
from electron EDM and $\dcp$ of $\bsg$. In addition, coupling strength  measurements of $h$ and heavy Higgs searches at the LHC
would also provide the complementary probes. In addition to these constraints, we also study potential signatures at the LHC.
We primarily focus on three searches at the 14 TeV LHC \bzh\footnote{Discussions on similar processes can also be  found in 
Refs.~\cite{Aaboud:2017cxo,Sirunyan:2019xls,Ferreira:2017bnx,Coyle:2018ydo}.}, \gga
and \bgba (for the discussion on latter two processes see also Refs.~\cite{ttbb,Gori:2016zto}). 
Induced by $\rbb$ the \bzh process can be searched at the LHC if $c_\gamma$ is nonzero and $m_A > m_h + m_Z$. 
On the other hand, $gg\to t \bar t  A \to t \bar t b \bar b$ process requires $\rtt$ and $\rbb$ both nonvanishing with mild dependence on $c_\gamma$. 
The final process \bgba also depends both on $\rbb$ and $\rtt$ but needs $m_A > 2 m_t$. Together with electron EDM and 
 $\dcp$ of $\bsg$, these processes can probe significant part of the parameter space for $\rho_{bb}$-EWBG.

Note that complex $\rtt$ can provide a more robust mechanism for 
EWBG~\cite{Fuyuto:2017ewj,Fuyuto:2019svr}. One may also have complementary probes
for the $\rho_{tt}$-EWBG from \gga and \bgba processes. Nonvanishing $\rho_{tt}$ motivates the conventional 
$gg\to H\to t \bar t$~\cite{Aaboud:2017hnm,Sirunyan:2019wph} (see also~\cite{Carena:2016npr})
search or $gg\to t \bar t A/H \to t \bar t t \bar t$~\cite{ttbb,Gori:2016zto,fourtop}, i.e., the four top search.
Though the former process suffers from large interference with the overwhelming QCD $gg\to t\bar t$ 
background~\cite{Carena:2016npr}, recent searches performed  by both 
ATLAS~\cite{Aaboud:2017hnm} and CMS~\cite{Sirunyan:2019wph} found some sensitivity.
When both $\rbb$ and $\rtt$ are nonzero,
one may also have $gg\to b \bar b A/H \to b \bar b t \bar t $, which are covered
in Refs.~\cite{ttbb,Gori:2016zto}.

\section{Parameter space}\label{param}
Let us find the allowed parameter space for $m_A$, $m_H$ and $m_{H^\pm}$.
The parameters in Eq.~\eqref{pot} are required to satisfy perturbativity, tree-level
unitarity and vacuum stability conditions, 
for which we utilized the public tool 2HDMC~\cite{Eriksson:2009ws}. 
We express the quartic couplings $\eta_1$, $\eta_{3{\rm -}6}$ in 
terms of $m_h$, $m_H$, $m_{H^\pm}$, $m_A$, $\mu_{22}$, $\gamma$, and $v$ as~\cite{Davidson:2005cw}
\begin{align}
& \eta_1 = \frac{m_h^2 s_\gamma^2 + m_H^2 c_\gamma^2}{v^2},\\
& \eta_3 =  \frac{2(m_{H^\pm}^2 - \mu_{22}^2)}{v^2},\\
& \eta_4 = \frac{m_h^2 c_\gamma^2 + m_H^2 s_\gamma^2 -2 m_{H^\pm}^2+m_A^2}{v^2},\\
& \eta_5 =  \frac{m_H^2 s_\gamma^2 + m_h^2 c_\gamma^2 - m_A^2}{v^2},\\
& \eta_6 =  \frac{(m_h^2 - m_H^2)(-s_\gamma)c_\gamma}{v^2}.
\end{align}
The quartic couplings $\eta_2$ and $\eta_7$ do not enter scalar masses, 
nor the mixing angle $\gamma$. 
Therefore in our analysis we take $v$, $m_h$, 
and $\gamma$, $m_A$, $m_H$, $m_{H^\pm}$, $\mu_{22}$, $\eta_2$, $\eta_7$ 
as the phenomenological parameters.
Further, to save computation time, we randomly generated these parameters 
in the following ranges:
$\mu_{22} \in [0, 1000]$  GeV,
$m_A \in [250, 600]$ GeV,
$m_H \in [250, 600]$ GeV, 
$m_{H^\pm} \in [250, 600]$ GeV, 
$\eta_2 \in [0, 6]$, $ \eta_7 \in [-6, 6]$, while satisfying $m_h = 125$ GeV with 
$c_\gamma = 0.1$\footnote{Note that, for successful $\rbb$ induced EWBG, one
requires non vanishing $c_\gamma$ as discussed in Ref.~\cite{Modak:2018csw}.
It was shown that for $c_\gamma \sim 0.1$, current data still allows $\irbb\sim$0.15--0.2, while $|\irbb|\gtrsim0.058$ is sufficient 
to account for the observed BAU~\cite{Modak:2018csw}.}.
The randomly generated parameters are then fed into 2HDMC for scanning. 2HDMC utilizes~\cite{Eriksson:2009ws} 
$m_{H^\pm}$ and $\Lambda_{1-7}$ as the input parameters in the Higgs basis whereas $v\simeq 246$ GeV. 
In order to match the 2HDMC convention, we identify $\eta_{1-7}$ as $\Lambda_{1-7}$ and, take $-\pi/2\leq \gamma \leq \pi/2$. 
For the positivity conditions of the Higgs potential of Eq.~\eqref{pot}, the parameter $\eta_2>0$ along with other more
involved conditions implemented in 2HDMC. We further conservatively demand $|\eta_i| \leq 6$.

\begin{figure*}[t]
\centering
\includegraphics[width=.45 \textwidth]{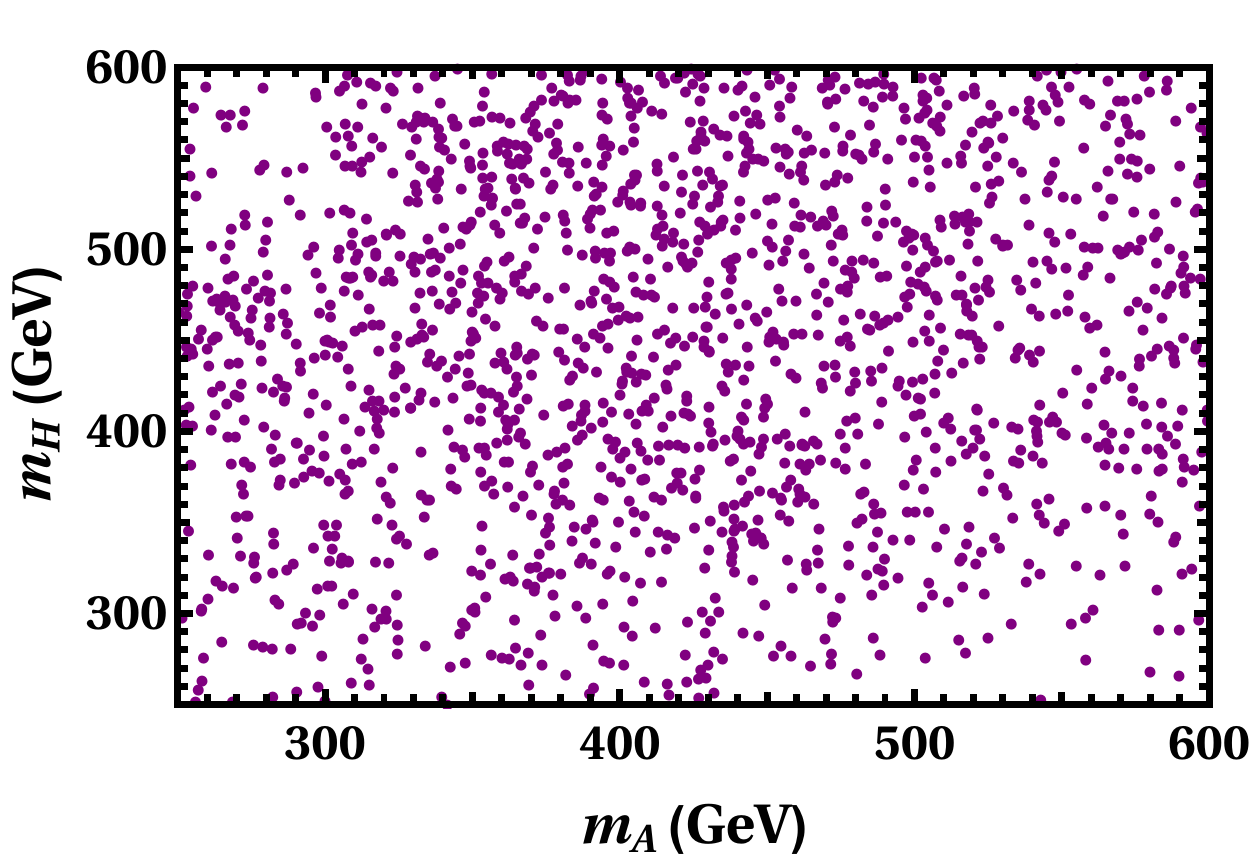}
\includegraphics[width=.45 \textwidth]{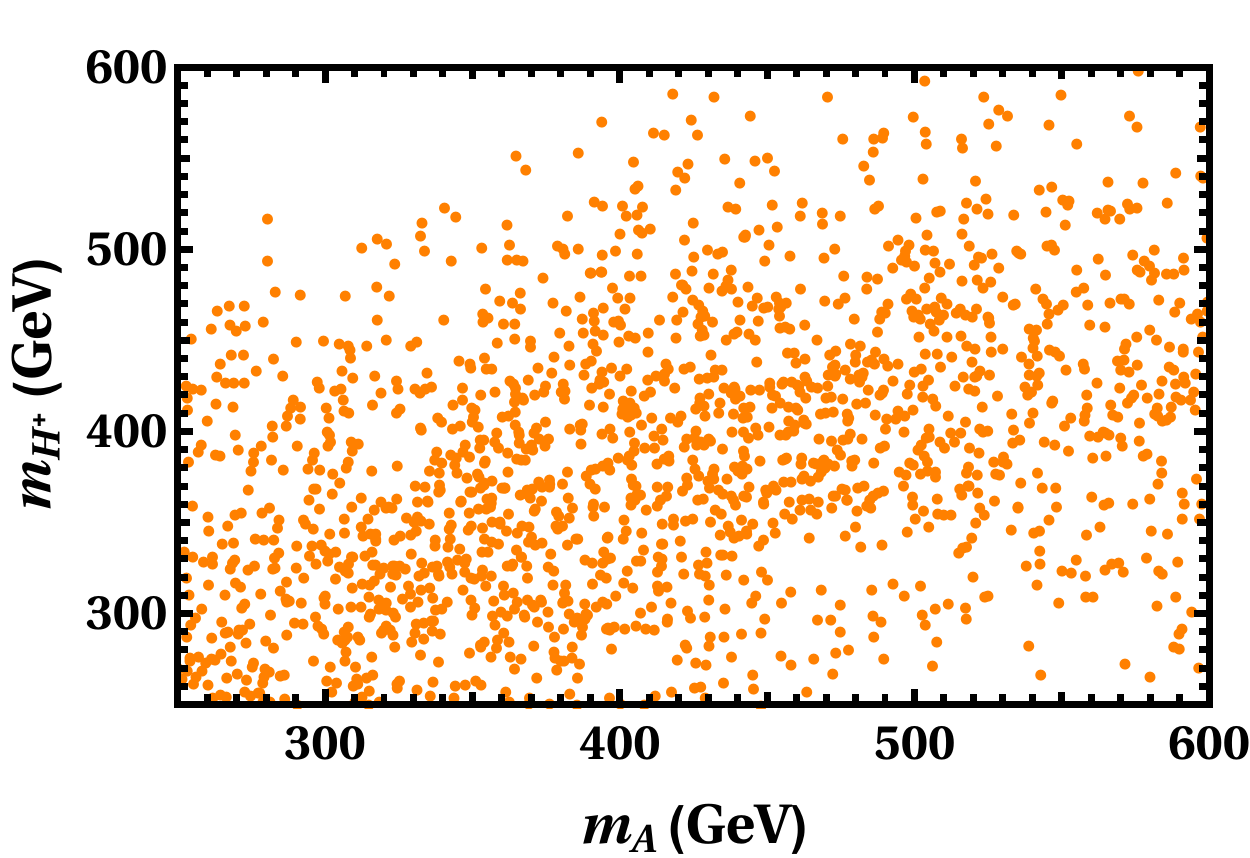}
\caption{Scanned points plotted in $m_A$--$m_H$ (left) and $m_A$--$m_{H^\pm}$ (right) plane. Here the scanned
points satisfy the tree level unitarity, perturbativity and positivity conditions as well as the $T$ parameter constraint.
However, not all points lead to the strong first-order EWPT.}
\label{scan}
\end{figure*}

Next we impose the stringent oblique $T$ parameter~\cite{Peskin:1991sw} constraint,
which restricts hierarchical structures among the scalar masses $m_H$, $m_A$ and $m_{H^\pm}$~\cite{Froggatt:1991qw,Haber:2015pua}, 
and hence $\eta_i$s. Utilizing the expression given in Ref.~\cite{Haber:2015pua} the points that
passed unitarity, perturbativity and positivity conditions from 2HDMC, are further required 
to satisfy the $T$ parameter constraint within the $2\sigma$ error~\cite{Baak:2014ora}.
These points are denoted as ``scanned points''.
We plot the scanned points in the $m_A$--$m_H$ and $m_A$--$m_{H^\pm}$ planes in the left and right panels of
Fig.~\ref{scan}, which illustrates that significant amounts of the allowed points exists.
A more detailed discussions on the scanning procedure can be found in Refs.~\cite{Hou:2019qqi,Hou:2019mve}. 
At this point, we have not yet required that EWPT should be strongly first order, and not all points are compatible with the $\rho_{bb}$ EWBG.

\begin{table*}[htbp!]
\centering
\begin{tabular}{c |c| c| c| c | c | c| c | c |c| c |c  c c| cc}
\hline
BP  & $\eta_1$ &  $\eta_2$   &  $\eta_3$   & $\eta_4$  & $\eta_5$ & $\eta_6$  & $\eta_7$ 
& $m_{H^\pm}$  & $m_A$ & $m_H$ &  $\frac{\mu_{22}^2}{v^2}$\\
 &&&&&&&& (GeV) & (GeV) & (GeV)&\\ 
\hline
&&&&&&&&&&&\\
$a$           & 0.282  & 2.034 & 4.053  & $-1.039$  & 1.343 & $-0.243$ & 1.231  & 391 & 285 & 405  & 0.5\\
$b$           & 0.289  & 1.959 & 4.064   & $-0.418$  & 1.56  & $-0.316$ & $-1.216$ & 414 & 334 & 456  & 0.8\\
$c$           & 0.303  & 0.413 & 5.129   & $-0.477$  & 1.534 & $-0.455$ & 0.457  & 508 & 444 & 541  & 1.7 \\ 
\hline
\hline
\end{tabular}
\caption{Parameter values of three benchmark points chosen from the scanned points in Fig.~\ref{scan}, which are consistent with the strong first-order EWPT.}
\label{bench}
\end{table*}

To find the constraints on $\rbb$ and $\rtt$ and, subsequently analyze the potential of
future probes we choose three benchmark points (BPs) from the scanned points in Fig.~\ref{scan},
which are summarized in Table~\ref{bench}. 
Here, we also demand that the chosen parameter sets give rise to the strong first-order EWPT.
The BP$a$ and BP$b$ are chosen such that $m_A < 2m_t$. Since there is no suppression from 
$\mathcal{B}(A\to t \bar t)$, such a choice would enhance the discovery potential of \bzh and \gga.
For the BP$c$, where $m_A > 2m_t$, the \bgba process\footnote{Note that for BP$a$ and BP$b$
one may have $bg\to bH \to b \bar b t \bar t$, which can resemble similar final state topologies as in
\bgba. This would be discussed in the Sec.~\ref{btt}.} can provide additional probes for the parameter space.
Further, for all three BPs, $A$ is assumed to be lighter than $H$ and $H^\pm$
to forbid $A\to Z H$ and $A \to H^\pm W^\mp$ decays and 
boost the discovery potential of these processes to some extent. Heavier $A$ are indeed possible, but 
the cross sections are reduced due to rapid fall in the parton luminosity.
We also remark that one requires~\cite{ewbg} sub-TeV $m_A$, $m_{H^\pm}$ and $m_H$
for the strong first-order EWPT, which is required for conventional sub-TeV
EWBG~\cite{ewbg_2hdm,ewbg_susy,ewbg_singletSM} (for high-scale EWBG, see, e.g., Refs.~\cite{highewbg}). 

In the following we will scrutinize the relevant constraints on $\rbb$ and $\rtt$. For simplicity,
we assume that $\rho_{ij}$ except for $\rbb$, $\rtt$ and $\rho_{ee}$ 
are negligibly small so as not to affect our main discussion. 
The impacts of nonzero $\rho_{ij}$ would be discussed later part of the paper.

\subsection{Flavor Constraints}
There exist several constraints from flavor physics that restricts the parameter space. 
In particular, the following three observables are relevant: (i) the branching ratio measurement of $B \to X_s \gamma$ ($\bsg$), (ii)
the asymmetry of the CP asymmetry between the charged and neutral $B\to X_s \gamma$ decays ($\dcp$)
and (iii) the $B_q$-$\overline{B}_q$ ($q=d,s$) mixings.

\begin{figure*}[htbp!]
\centering
\includegraphics[width=.316 \textwidth]{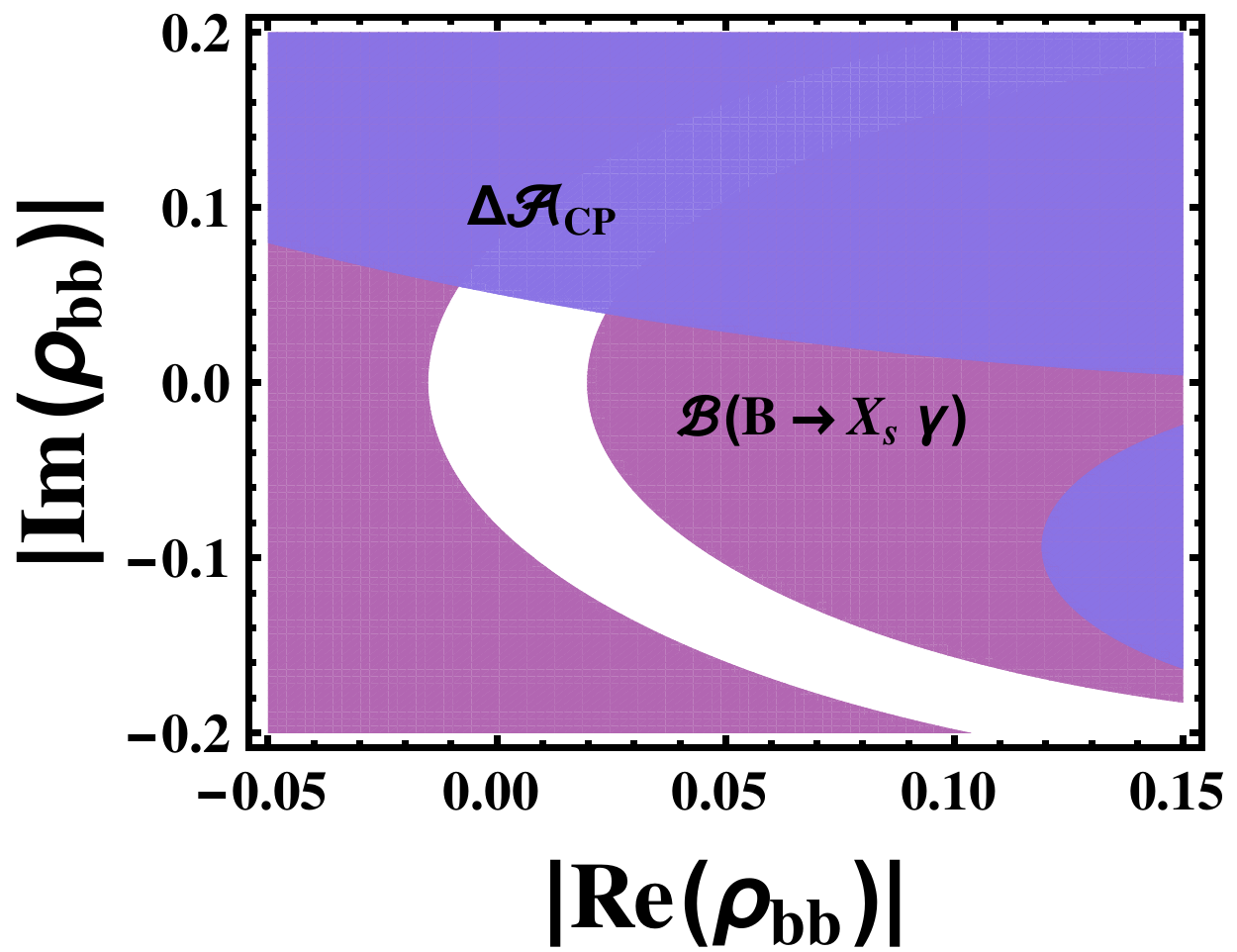}
\includegraphics[width=.33 \textwidth]{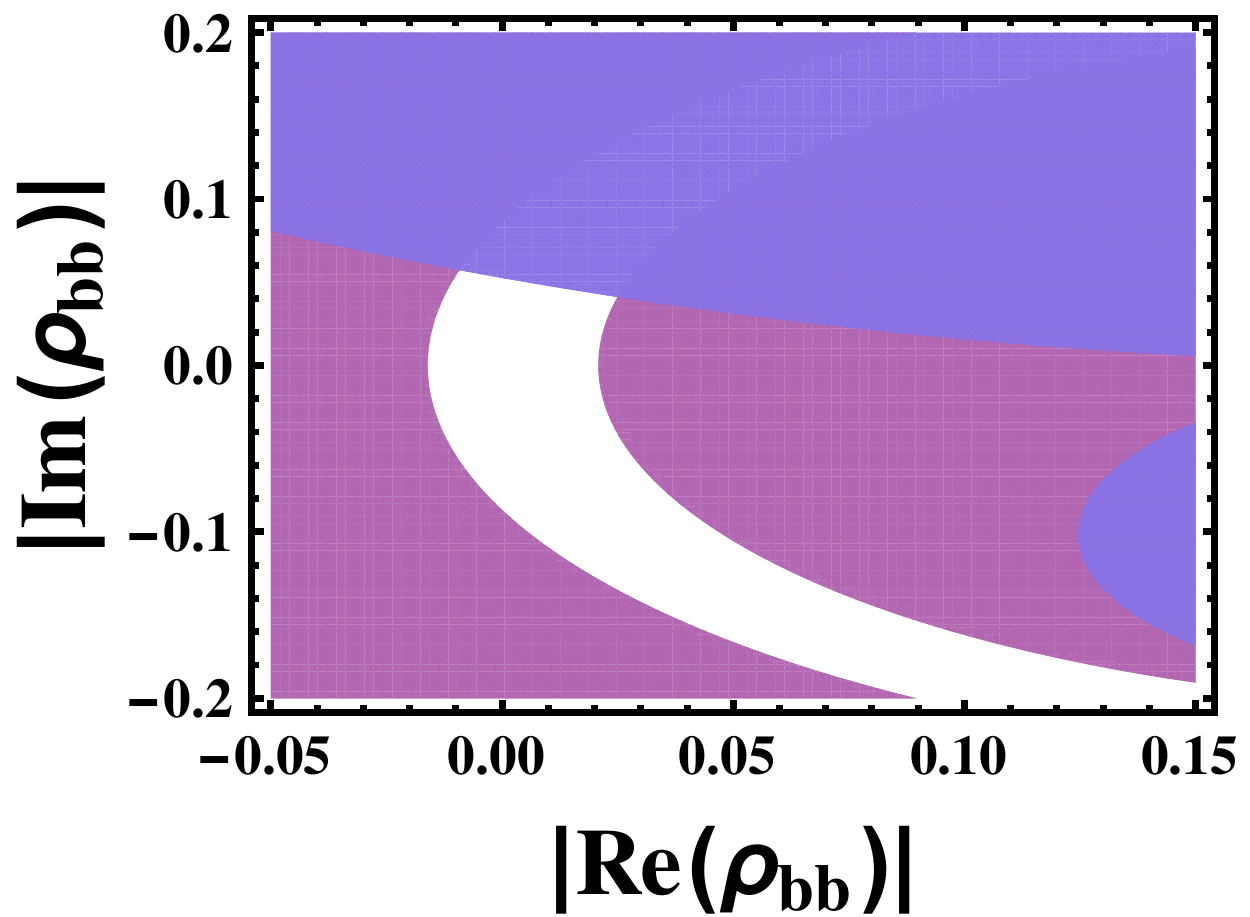}
\includegraphics[width=.33 \textwidth]{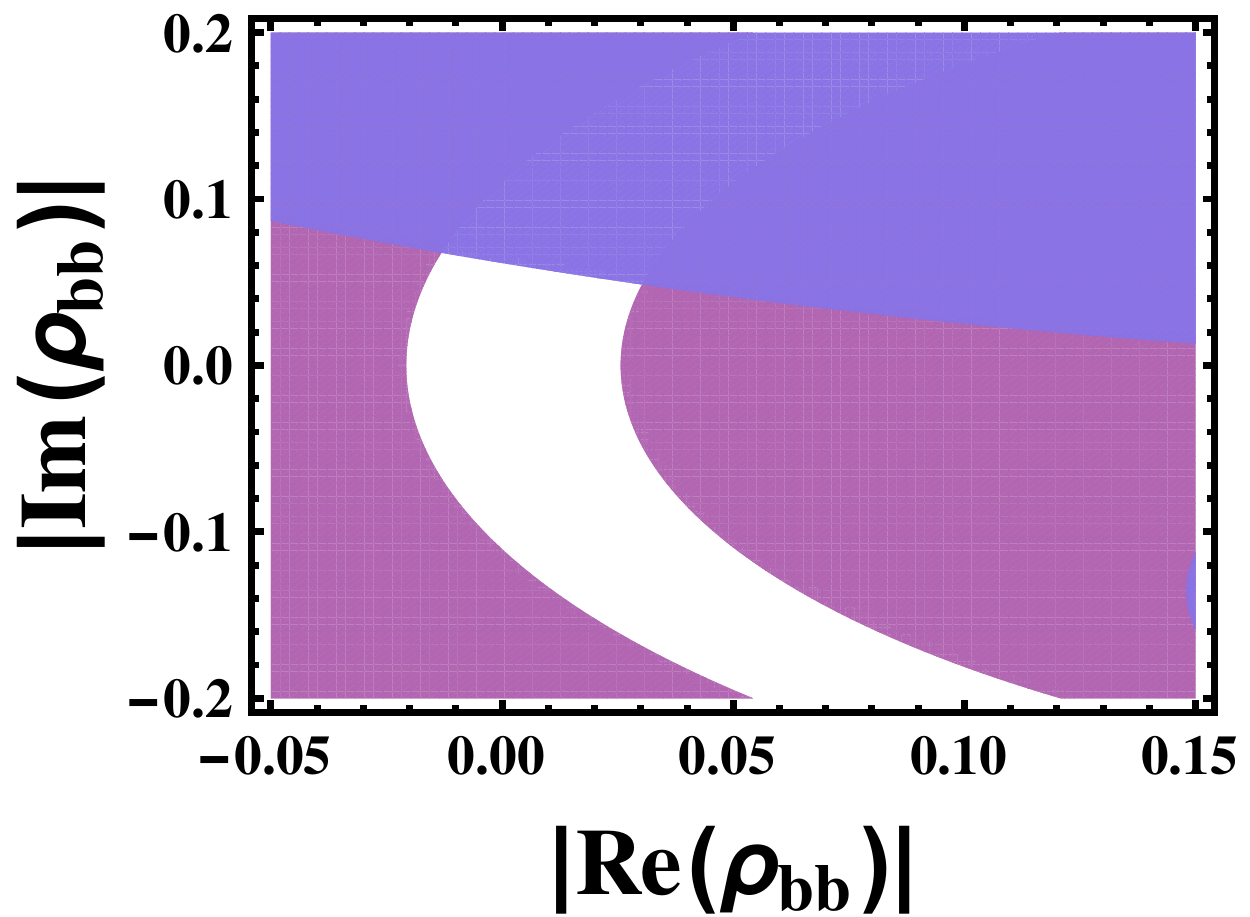}
\caption{The constraints on $\rbb$ from $\bsg$ (purple) and $\dcp$ (blue) measurements for the BP$a$ (left), 
BP$b$ (middle) and BP$b$ (right) respectively. All three figures are generated assuming $\rtt = 0.5$. See text for details.}
\label{scanim}
\end{figure*}

Let us first focus on $\bsg$. Non-zero $\rbb$ and $\rtt$ modify $\bsg$ via top quark and charged Higgs boson loop.
The modification is parametrized by the (LO) Wilson coefficients $C^{(0)}_{7,8}$
at the matching scale $\mu = m_W$
\begin{align}
C^{(0)}_{7,8}(m_W)= F^{(1)}_{7,8}(x_t)+\delta C_{7,8}^{(0)}(\mu_W),\label{c78}
\end{align}
where, $\overline{m}_t(m_W)$ is the top quark $\overline{\mbox{MS}}$ running mass at the $m_W$ scale
with $x_t=(\overline{m}_t(m_W)/m_W)^2$. The expression for $F^{(1)}_{7,8}(x)$ can 
be found in the Refs.~\cite{Ciuchini:1997xe,Chetyrkin:1996vx}, whereas $\delta C_{7,8}^{(0)}(\mu_W)$ 
the LO (leading order) charged Higgs contributions.
At LO, $\delta C_{7,8}^{(0)}(\mu_W)$ is expressed as~\cite{Altunkaynak:2015twa}
\begin{align}
 \delta C_{7,8}^{(0)}(m_W)\simeq &\frac{|\rtt|^2}{3\lambda_t^2}F^{(1)}_{7,8}(y_{H^+}) 
 -\frac{\rtt\rho_{bb}}{\lambda_t\lambda_b}F^{(2)}_{7,8}(y_{H^+}),
\end{align}
with $y_{H^+}=(\overline{m}_t(m_W)/m_{H^+})^2$ while, 
the full expression for $F^{(2)}_{7,8}(y_{H^+})$ can be found in Ref.~\cite{Ciuchini:1997xe}.
The current world average of $\bsg_{\text{exp}}$ extrapolated to the photon-energy cut $E_0=1.6$ GeV 
is found by the HFLAV Collaboration to be $(3.32\pm0.15)\times 10^{-4}$~\cite{Amhis:2016xyh}. The
next-to-next-to LO (NNLO) $\bsg$ prediction in the SM for the same photon-energy cut 
is $(3.36\pm0.23)\times 10^{-4}$~\cite{Czakon:2015exa}. In order to find the constraint,
we adopt the prescription outlined in Ref.~\cite{Crivellin:2013wna} and define
\begin{align}
 R_{\text{exp}}=\frac{\bsg_{\text{exp}}}{\bsg_{\text{SM}}}.
\end{align}
Based on our LO Wilson coefficients, we further express
\begin{align}
 R_{\text{theory}}=\frac{\bsg_{\text{g2HDM}}}{\bsg_{\text{SM}}},
\end{align}
and take $m_W$ and $\overline{m}_b(m_b)$ respectively as the matching scale and the low-energy scales.
Finally, we demand $R_{\text{theory}}$ to remain within the $2\sigma$ error of $R_{\text{exp}}$. 
In Fig.~\ref{scanim} the excluded regions are shown as the purple shaded regions in 
the $\rrbb$--$\irbb$ plane for three BPs. Here, we assume $\rtt = 0.5$. Flavor 
constraints on $\rho_{tt}$ is moderately strong, with $B_{d,s}$-$\overline{B}_{d,s}$ mixings providing the most stringent constraint 
on $\rho_{tt}$ for $500 \lesssim m_{H^+} \lesssim 650$ GeV, which is the 
ballpark mass ranges of $m_{H^\pm}$ for all the three BPs. 
The $B_q$-$\overline{B}_q$ mixing amplitude $M^q_{12}$ receives modification
from the charged Higgs and $W$ bosons loop with $t$ quark.
Utilizing the expression for $B_q$-$\overline{B}_q$ mixing in
type-II 2HDM~\cite{Geng:1988bq}, it is found in Ref.~\cite{Altunkaynak:2015twa} that
\begin{align}
 \frac{M^q_{12}}{M^{q\;\rm{SM}}_{12}} = 1+ \frac{I_{WH}(y_W, y_H, x) + I_{HH}(y_H)}{I_{WW}(y_W)},\label{bqmixing}
\end{align}
where $y_i = m_t^2/m_i^2$ ($i = W, H^\pm$) and $x = m_{H^\pm}^2/m_W^2$ with $m_t$ and $m_W$
being the masses of the top quark and $W$ bosons. The expressions for 
$I_{WW}$, $I_{WH}$ and $I_{HH}$ are respectively given by~\cite{Altunkaynak:2015twa}
\begin{align}
 I_{WW}&=1+\frac{9}{1-y_W}-\frac{6}{(1-y_W)^2}-\frac{6}{y_W}\left(\frac{y_W}{1-y_W}\right)^3 \ln y_W,\\
 I_{WH}&\simeq \left(\frac{\rho_{tt}^*}{\lambda_t}+\frac{V_{cb}\rho_{ct}^*}{V_{tb}\lambda_t}\right)
 \left(\frac{\rho_{tt}}{\lambda_t}+\frac{V_{cq}^*\rho_{ct}}{V_{tq}^*\lambda_t}\right) y_H\nonumber\\
 &\quad \times \bigg[ \frac{(2x-8) \ln y_H}{(1-x) (1-y_H)^2} + 
 \frac{6x \ln y_W}{(1-x)(1-y_W)^2}- \frac{8- 2 y_W}{(1-y_W)(1-y_H)}\bigg] ,\label{IWHexpr}\\
 I_{HH}&\simeq \left(\frac{\rho_{tt}^*}{\lambda_t}+\frac{V_{cb}\rho_{ct}^*}{V_{tb}\lambda_t}\right)^2
 \left(\frac{\rho_{tt}}{\lambda_t}+\frac{V_{cq}^*\rho_{ct}}{V_{tq}^*\lambda_t}\right)^2\bigg( \frac{ 1+ y_H}{(1- y_H)^2} 
 + \frac{2 y_H \ln y_H}{(1- y_H)^3}\bigg)y_H .\label{IHHexpr}
\end{align}
For $|\rho_{tt}|\sim\mathcal{O}(1)$ coupling $\rho_{ct}$ is strongly constrained due
to $|V_{cq}/ V_{tq}|\sim 25$ ($q$ = $d$, $s$) enhancement~\cite{Altunkaynak:2015twa},
as can be seen from Eqs.\eqref{IWHexpr}  and \eqref{IHHexpr}. As we are primarily interested in the parameter 
space where $\rho_{tt}$ is $\mathcal{O}(1)$, we turn off $\rho_{ct}$ throughout our paper for simplicity.
The 2018 summer results of UTfit finds~\cite{utfitrse}:
\begin{align}
 &C_{B_d}\in 1.05\pm 0.11 ,\nn\\
 &C_{B_s}\in 1.110\pm0.090 ,\nn\\
 &\phi_{B_d}\in -2.0\pm1.8~~[\mbox{in}~^{\circ} ],\nn\\
 &\phi_{B_s}\in 0.42\pm0.89~~[\mbox{in}~^{\circ} ].
\end{align}
with $M^q_{12}/M^{q\;\rm{SM}} = C_{B_q} e^{2i \phi_{B_q}}$.
Under the assumption on the $\rho^F_{ij}$ couplings made in our analysis, we have $M^q_{12}/M^{q\;\rm{SM}}= C_{B_q}$. 
Allowing $2\sigma$ errors on $C_{B_d}$ and $C_{B_s}$ we find that 
$B_{s,d}$-$\overline{B}_{d,s}$ mixings exclude $|\rtt| \gtrsim 0.9$ for BP$a$ and BP$b$ and, $|\rtt| \gtrsim 1$ for BP$c$.

One of the most powerful probes of $\irbb$ is the direct CP asymmetry $\mathcal{A}_{\text{CP}}$~\cite{Kagan:1998bh} of
$B\to X_s \gamma$. It is advocated in Ref.~\cite{Benzke:2010tq}, however, that $\dcp$ is even more sensitive to the CP-violating
couplings, which is defined as~\cite{Benzke:2010tq}
\begin{align}
 \dcp = \mathcal{A}_{B^-\to X_s^- \gamma} - \mathcal{A}_{B^0\to X_s^0 \gamma}
 \approx 4 \pi^2 \alpha_s \frac{\tilde{\Lambda}_{78}}{m_b}\mbox{Im}\bigg(\frac{C_8}{C_7}\bigg),\label{acp}
\end{align}
where $\tilde{\Lambda}_{78}$ and $\alpha_s$ denote a hadronic parameter and the strong coupling constant at $\overline{m}_b(m_b)$ scale, respectively.
One expects that $\tilde{\Lambda}_{78}$ has a similar scale of $\Lambda_{\text{QCD}}$.
In Ref.~\cite{Benzke:2010tq}, it is found that $17~\text{MeV}\le \tilde{\Lambda}_{78}\le 190~\text{MeV}$.
On the other hand, recently Belle measured $\dcp=(+3.69\pm2.65\pm0.76)\%$~\cite{Watanuki:2018xxg}, 
where the first uncertainty is statistical while the second one is systematic.
Allowing $2\sigma$ error on the Belle measurement, we show the regions excluded by $\dcp$ 
in blue shade in Figs.~\ref{scanim} for the three BPs. 
Here, we choose the average value of $\tilde{\Lambda}_{78}$ i.e., 89 MeV for illustration.
We stress that the constraint shown in Figs.~\ref{scanim} depends heavily on the value 
of $\tilde{\Lambda}_{78}$. The larger $\tilde{\Lambda}_{78}$ would make the constraint stronger.
We also remark that we utilize the LO Wilson coefficients in Eq.~\eqref{c78} 
as a first approximation for simplicity. Note that the excluded regions by $\dcp$ measurement 
in  Fig.~\ref{scanim} is asymmetric and constrains positive $\irbb$ more stringently. This
is solely due to our choice of $\rtt = 0.5$. If we take $\rtt = -0.5$,
the blue shaded regions would flip and exclude the negative regions of $\irbb$.

We note in passing that if $\rho_{tt}$ is also complex, $\dcp$ can be zero if the 
complex phases of $\rho_{tt}$ and $\rho_{bb}$ are aligned, i.e., $\text{Im}(\rho_{tt}\rho_{bb})$=0,
equivalently, $\text{Re}\rho_{bb}/\text{Re}\rho_{tt}=-\text{Im}\rho_{bb}/\text{Im}\rho_{tt}$. Such a 
phase alignment is discussed in Ref.~\cite{Fuyuto:2019svr}.

\subsection{EDMs}\label{sec:edms}

\begin{figure}[t]
\center
\includegraphics[width=6cm]{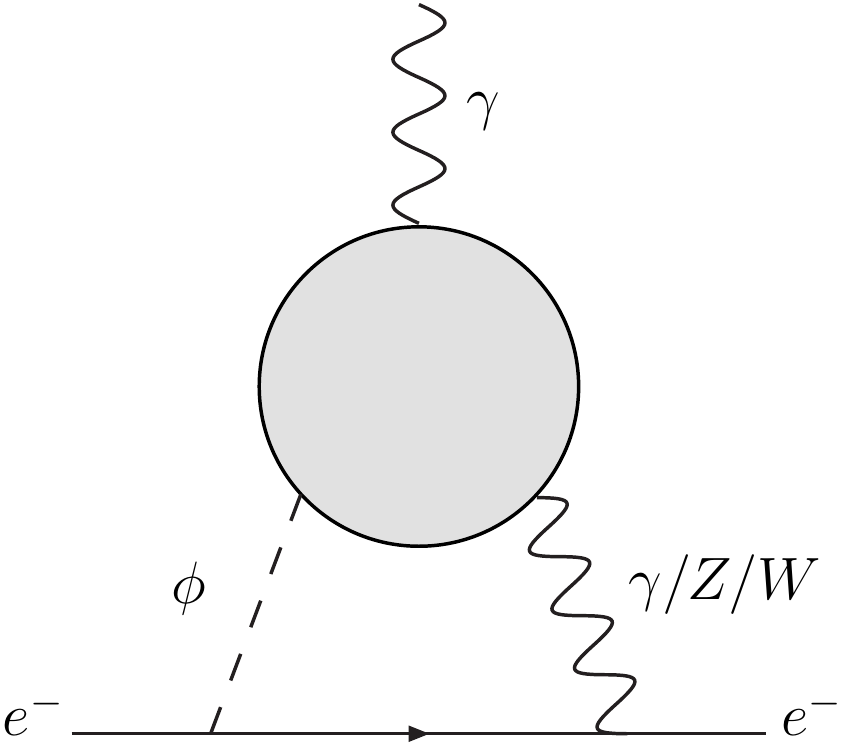}
\caption{Two-loop Barr-Zee diagrams contributing to the electron EDM, where $\phi=h,H,A,H^\pm$. 
The shaded loop collectively represents the scalar, fermion and gauge boson loops. The total contribution 
is given by their sum, $d_e = d_e^{\phi\gamma}+d_e^{\phi Z}+d_e^{\phi W}$.}
\label{fig:de}
\end{figure}

The complex phase of $\rho_{bb}$ is severely constrained by EDMs of the electron, neutron, and atoms, etc. 
Currently, the most stringent experimental bound comes from EDM of thorium monoxide (ThO), which is approximately given by 
\begin{align}
d_{\text{ThO}} = d_e + \alpha_{\text{ThO}}C_S,
\end{align} 
where $d_e$ is the electron EDM and $C_S$ is the coefficient of the nuclear
spin-independent interaction (NSID), which are respectively defined as
\begin{align}
\mathcal{L}_{\text{EDM}} = -\frac{i}{2}d_eF^{\mu\nu}\bar{e}\sigma_{\mu\nu}\gamma_5e,\quad
\mathcal{L}_{eN}^{\text{NSID}} = -\frac{G_F}{\sqrt{2}}C_S(\bar{N}N)(\bar{e}i\gamma_5e),
\end{align}
where $F^{\mu\nu}$ denotes the field strength tensor of electromagnetism and $G_F$ is the Fermi coupling constant.
The coefficient $\alpha_{\text{ThO}}$ 
is estimated as $\alpha_{\text{ThO}}=1.5\times10^{-20}$~\cite{Fuyuto:2018scm}. 
The latest experimental value of $d_{\text{ThO}}$ is placed by ACME Collaboration in 2018 (ACME18) as
\begin{align}
d_{\text{ThO}} = (4.3\pm4.0)\times 10^{-30}~e~\text{cm},\label{dThO_acme18}
\end{align}
from which under the assumption of $C_S=0$ the electron EDM has an upper bound of
\begin{align}
|d_e|<1.1\times 10^{-29}~e~\text{cm}.\label{de_acme18}
\end{align}
In our scenario, $d_e$ is predominantly induced by two-loop Barr-Zee diagrams
as depicted in Fig.~\ref{fig:de}, which are decomposed into the three parts:
\begin{align}
d_e = d_e^{\phi\gamma}+d_e^{\phi Z}+d_e^{\phi W},
\end{align}
where $\phi=h, H, A$ for the first two terms and $\phi=H^\pm$ for the last term. 
Let us denote the contribution of $i$-species to $d_e^{\phi\gamma}$ as $(d_e^{\phi\gamma})_i$. 
If $\rho_{bb}$ is the only source of $CP$ violation, $d_e\simeq (d_e^{\phi\gamma})_b$. With $\text{Im}\rho_{bb}$ 
required by $\rho_{bb}$-EWBG mechanism, $d_e$ is so large that one cannot avoid the ACME18 bound as noted in 
Ref.~\cite{Modak:2018csw}. This fact suggests two options: (i) the alignment limit ($c_\gamma\to0$) and (ii)
cancellation mechanism. As discussed in Ref.~\cite{Fuyuto:2019svr}, however, the first option may not be consistent 
with EWBG in g2HDM since the BAU would be suppressed with decreasing $c_\gamma$. We thus consider the second option. 
Even though we identify the parameter space for the cancellation in Ref.~\cite{Modak:2018csw}, we do not show its detail there, 
and moreover, $d_e^{\phi W}$, which can come into play in the cancellation region, is missing. We therefore update our previous 
analysis taking all the relevant contributions into consideration.

If there exist more than two $CP$-violating phases, we could tune the parameters in such a way that $d_e$ 
becomes small. While it is nothing more than the parameter turning, we still classify the cancellation parameter 
space into two kind. We call a cancellation \textit{structured cancellation} if it happens when the hierarchical 
structures of the $\rho_{ij}$ matrices closely resemble those of the SM Yukawa matrices, and anything else is 
\textit{unstructured cancellation}. It is revealed in Ref.~\cite{Fuyuto:2019svr} that the parameter space of $\rho_{tt}$-EWBG 
accommodates the structured cancellation. We here scrutinize the type of the  cancellation in $\rho_{bb}$-EWBG.

Following a method adopted in Ref.~\cite{Fuyuto:2019svr}, we split $(d_e^{\phi\gamma})_b$ into two parts as\footnote{ By 
convention in this paper, the sign of $\gamma$ is opposite to that in Ref.~\cite{Fuyuto:2019svr}.}
\begin{align}
(d_e^{\phi\gamma})_f = (d_e^{\phi\gamma})_f^\text{mix}+(d_e^{\phi\gamma})_f^\text{extr},
\end{align}
where
\begin{align}
\frac{(d_{e}^{\phi\gamma})_f^{\text{mix}}}{e}
&=-\frac{3\alpha_{\rm em}Q_f^2s_{2\gamma}}{16\sqrt{2}\pi^3v}
\bigg[
	\text{Im}(\rho_{ee})\Delta f_f
	+\frac{\lambda_e}{\lambda_f}\text{Im}(\rho_{ff})\Delta g_f
\bigg], \\
\frac{(d_{e}^{\phi\gamma})_f^{\text{extr}}}{e}
&=\frac{3\alpha_{\text{em}}Q_f^2}{16\pi^3m_f}
\bigg[
	\text{Im}(\rho_{ee})\text{Re}(\rho_{ff})
	\Big\{c_\gamma^2f(\tau_{fh})+s_\gamma^2f(\tau_{fH})\pm g(\tau_{fA}) \Big\} \nonumber\\
&\hspace{2cm}
	+\text{Im}(\rho_{ff})\text{Re}(\rho_{ee})
	\Big\{c_\gamma^2g(\tau_{fh})+s_\gamma^2g(\tau_{fH})\pm f(\tau_{fA}) \Big\}
\bigg],\label{de_ex}
\end{align}
with $\alpha_{\text{em}}$ and $Q_f$ representing the fine structure constant and electric charges of $f$, respectively, and
$\Delta f_f = f(\tau_{fh})-f(\tau_{fH})$ and $\Delta g_f = g(\tau_{fh})-g(\tau_{fH})$ with
$\tau_{ij}=m_i^2/m_j^2$. $f(\tau)$ and $g(\tau)$ are the loop functions
and their explicit forms are shown in Appendix ~\ref{app:edm}. In our notation, the sign of $e$ is positive.
In the wave parenthesis in Eq.~(\ref{de_ex}), the upper sign is for up-type fermions and the lower is for down-type fermions, respectively.
For $c_\gamma\ll 1$ and $m_H\simeq m_A$, $(d_{e}^{\phi\gamma})_{t,b}^{\text{extr}}$ are approximated as
\begin{align}
\frac{(d_{e}^{H\gamma})_t^{\text{extr}}}{e}\simeq\frac{\alpha_{\text{em}}}{12\pi^3m_t}\text{Im}(\rho_{ee}\rho_{tt})\Big[f(\tau_{tH})+g(\tau_{tH})\Big], \\
\frac{(d_{e}^{H\gamma})_b^{\text{extr}}}{e}\simeq\frac{\alpha_{\text{em}}}{48\pi^3m_b}\text{Im}(\rho_{ee}\rho_{bb}^*)\Big[f(\tau_{bH})-g(\tau_{bH})\Big].
\end{align}
In the $\rho_{bb}$-EWBG scenario, $\text{Im}(\rho_{ee}\rho_{tt})=\rho_{tt}\text{Im}(\rho_{ee})$.
To make our discussion on the cancellation mechanism simpler, we consider a case in which 
$\text{Im}(\rho_{ee}\rho_{bb}^*)\simeq 0$ so that $(d_e^{\phi\gamma})_b \simeq (d_e^{\phi\gamma})_b^\text{mix}$. 
When $\rho_{ee}$ is nonzero, the primary contribution could be $(d_e^{\phi\gamma})_W$ as inferred from the fact 
that the $\phi$-$\gamma$-$\gamma$ vertex in $d_e^{\phi\gamma}$ is more or less common to the $h\to 2\gamma$ decay. 
Noting that the $W$-loop has only the ``mix" contribution since the Higgs couplings
to the $W$ bosons are the gauge couplings, one may find~\cite{Abe:2013qla}
\begin{align}
\frac{(d_{e}^{H\gamma})_W}{e}=\frac{(d_{e}^{H\gamma})_W^{\text{mix}}}{e}
=\frac{\alpha_{\text{em}}s_{2\gamma}}{64\sqrt{2}\pi^3v}\text{Im}(\rho_{ee})\Delta\mathcal{J}_W^\gamma,
\end{align}
where $\Delta\mathcal{J}_W^\gamma = \mathcal{J}^\gamma_W(m_h)-\mathcal{J}^\gamma_W(m_H)$
(for explicit form of $\mathcal{J}_W^\gamma$, see Appendix~\ref{app:edm}).
From the condition of $(d_e^{\phi\gamma})_t+(d_e^{\phi\gamma})_b+(d_e^{\phi\gamma})_W=0$, it follows that
\begin{align}
\frac{\text{Im}(\rho_{ee})}{\text{Im}(\rho_{bb})}
= -\frac{s_{2\gamma}\Delta g_b/4}{s_{2\gamma}[\Delta f_t+\Delta f_b/4-(3/16)\Delta \mathcal{J}_W^\gamma]+2\rho_{tt}[f(\tau_{tH})+g(\tau_{tH})]/\lambda_t}
\equiv -c\times\frac{\lambda_e}{\lambda_b}.\label{cancel}
\end{align}
It is found that $c=1.0\times 10^{-3}$ for $c_\gamma=0.1$, $\rho_{tt}=0.5$, $m_h=125$ GeV and $m_H=405$ GeV.
Therefore, the cancellation is possible but unstructured since $c$ deviates much from the unity as opposed to the $\rho_{tt}$-EWBG 
scenario~\cite{Fuyuto:2019svr}. Once this accidental cancellation happens, other contributions could become relevant. 
On the grounds of dimensional analysis, one can find that $d_e^{\phi Z}$ is suppressed by the $Z$ boson 
coupling to the electron, $g_{Zee}^{}=1/4-\sin\theta_W\simeq 0.02$ with $\theta_W$ representing the weak 
mixing angle, while $d_e^{\phi W}$ is not and becomes leading contribution.
The dominant contribution in $d_e^{\phi W}$ comes from the diagrams involving the top and bottom loops, 
which amounts to~\cite{BowserChao:1997bb,Fuyuto:2019svr}
\begin{align}
\frac{(d_e^{\phi W})_{t/b}}{e} 
& \simeq \frac{3\alpha_{\text{em}}|V_{tb}|^2}{128\pi^3s_W^2}\frac{m_t}{m_{H^\pm}^2}\text{Im}(\rho_{ee}\rho_{tt})J_1(\tau_{WH^\pm}, \tau_{tH^\pm}),
\end{align}
where $m_b=0$ and $V_{tb}$ is the (33) element of the CKM matrix, which is close to one~\cite{Tanabashi:2018oca}.
$J_1$ is the loop function listed in Appendix~\ref{app:edm}.
In general, this contribution has the $\rho_{bb}$ dependence but vanishes in the case of $m_b=0$.
Note that $(d_e^{\phi W})_{t/b}$ is absent in the softly-broken $Z_2$ 2HDMs.
For one of the $\rho_{bb}$-EWBG parameter points, e.g., $\text{Im}(\rho_{bb})=-0.15$,
one would get $(d_e^{\phi W})_{t/b}\simeq1.2\times10^{-29}~e~\text{cm}$ in the cancellation
region specified by Eq.~(\ref{cancel}), together with $m_{H^\pm}=391$ GeV and $\rho_{tt}=0.5$, 
which slightly exceeds the ACME18 bound. Therefore, the allowed region is not exactly determined by
the cancellation condition but it occurs in its vicinity, as we show in our numerical analysis conducted below.
It should be noted that even though $\rho_{tt}$ is real
in $\rho_{bb}$-EWBG, its magnitude can be constrained by the electron EDM due to the proportionality of $\rho_{tt}\text{Im}(\rho_{ee})$.

Now we move on to discuss the $C_S$ contribution. We estimate $C_S$ 
using the CP-violating 4-fermion interactions between the quarks and electron defined as
\begin{align}
\mathcal{L}_{4f}^{\text{CPV}} = \sum_qC_{qe}(\bar{q}q)(\bar{e}i\gamma_5 e),
\end{align}
where $C_{qe} = \sum_{\phi=h,H,A}g^S_{\phi\bar{q}q}g^P_{\phi\bar{e}e}/m_\phi^2$ (explicit 
forms of $g^S_{\phi\bar{q}q}$ and $g^P_{\phi\bar{e}e}$ are shown in Appendix~\ref{app:edm})
With those, $C_S$ is estimated as~\cite{Dekens:2018bci}
\begin{align}
C_S = -2v^2
\bigg[
	6.3(C_{ue}+C_{de})+C_{se}\frac{41~\text{MeV}}{m_s}+C_{ce}\frac{79~\text{MeV}}{m_c}
	+62~\text{MeV}
	\left(\frac{C_{be}}{m_b}+\frac{C_{te}}{m_t}\right)
\bigg].
\end{align}
Note that for $c_\gamma\ll1$ and $m_H\simeq m_A$, $C_{qe}$ for up- and down-type quarks are, 
respectively, cast into the form~\cite{Fuyuto:2019svr}
\begin{align}
C_{ue} \simeq \frac{1}{2m_H^2}\text{Im}(\rho_{ee}\rho_{uu}),\quad
C_{de} \simeq \frac{1}{2m_H^2}\text{Im}(\rho_{ee}\rho_{dd}^*).
\end{align}
Therefore, the dependences of the CP-violating phases are the same as those of 
$(d_e^{\phi\gamma})^{\text{extr}}_u$ and $(d_e^{\phi\gamma})^{\text{extr}}_d$, respectively.

\begin{figure}[t]
\center
\includegraphics[width=10cm]{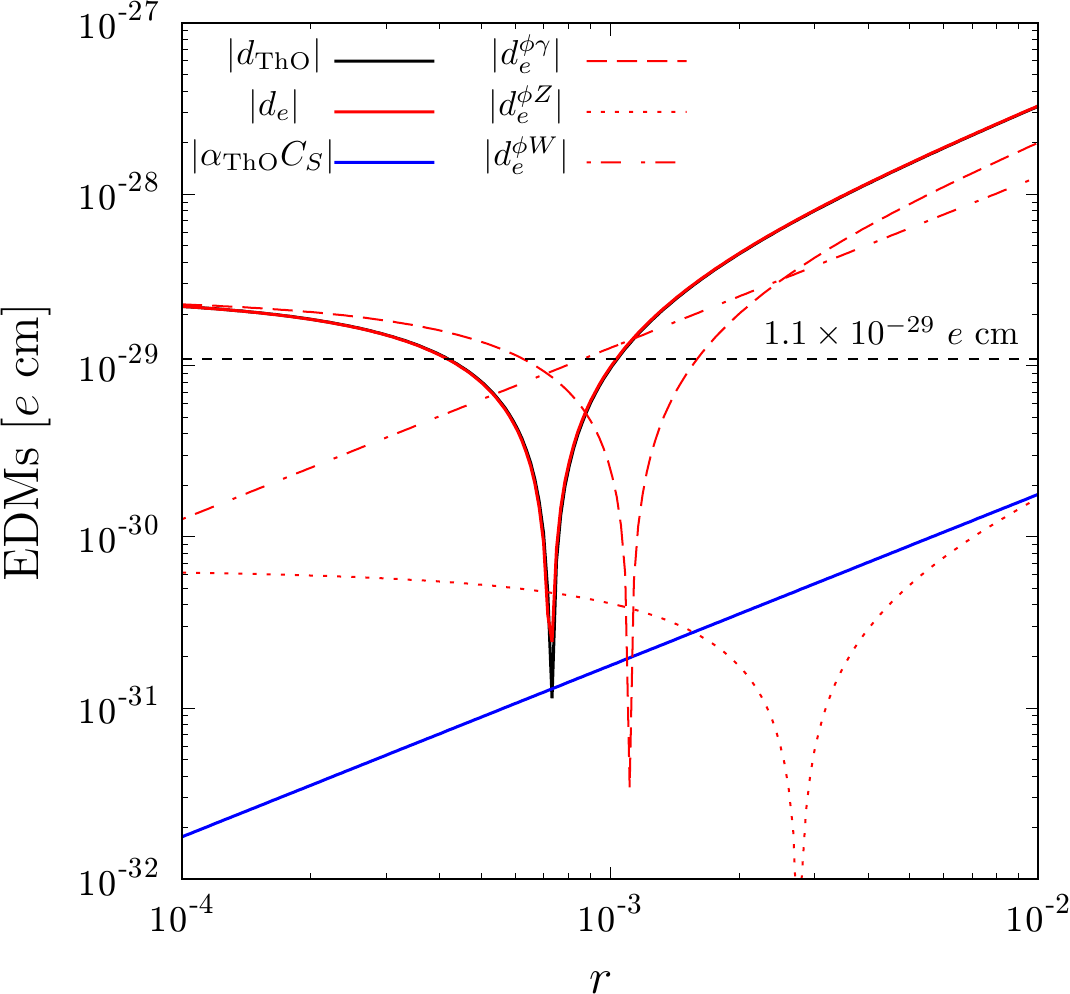}
\caption{Details of EDMs as functions of $r$. We take BP$a$ for the Higgs spectrum and
set $c_\gamma=0.1$, $\text{Re}(\rho_{bb})=0$, $\text{Im}(\rho_{bb})=-0.15$ and $\rho_{tt}=0.5$
as an example of the $\rho_{bb}$-EWBG scenario. Other $\rho_{ff}$ are fixed by 
$\text{Re}(\rho_{ee})=-r(\lambda_e/\lambda_b)\text{Re}(\rho_{bb})$ and $\text{Im}(\rho_{ee})=-r(\lambda_e/\lambda_b)\text{Im}(\rho_{bb})$. 
The ACME18 bound ($|d_e|<1.1\times 10^{-29}~e~\text{cm}$) is shown by the horizontal dotted line in black.}
\label{fig:dThO_r}
\end{figure}

In our numerical analysis, we parametrize $\rho_{ff}$, except for $\rho_{tt}$, as 
$\text{Re}(\rho_{ee})=-r(\lambda_e/\lambda_b)\text{Re}(\rho_{bb})$ and $\text{Im}(\rho_{ee})=-r(\lambda_e/\lambda_b)\text{Im}(\rho_{bb})$.
Though the CP-violating phases in the first and second generations of $\rho^F$ matrices 
have nothing to do with $\rho_{bb}$-EWBG, we fix them through the above relations. However, 
the effects of the extra CP violation are too small to affect our cancellation mechanism in $\rho_{bb}$-EWBG. 

In Fig.~\ref{fig:dThO_r}, $|d_{\text{ThO}}|$ and its details are shown as functions of $r$. 
We take BP$a$ for the Higgs spectrum and set $c_\gamma=0.1$, $\text{Re}(\rho_{bb})=0$,
$\text{Im}(\rho_{bb})=-0.15$ and $\rho_{tt}=0.5$ as an example of the $\rho_{bb}$-EWBG scenario. 
As seen, the magnitude of $\alpha_{\text{ThO}}C_S$ is much smaller than that of $d_e$, 
we thus can use the ACME18 bound of $|d_e|<1.1\times 10^{-29}~e~\text{cm}$, which is represented by the
horizontal dotted line in black, to constrain the parameter space. 
As discussed in Eq.~(\ref{cancel}), the cancellation happens in $d_e^{\phi\gamma}$ 
at around $r\simeq 1\times10^{-3}$, which is the consequences of 
$(d_e^{\phi\gamma})_t+(d_e^{\phi\gamma})_b+(d_e^{\phi\gamma})_W\simeq 0$. At this point, 
$d_e^{\phi W}$ becomes dominant and $|d_e|$ exceeds the ACME 18 bound. Nevertheless, the 
cancellation is still at work at around $r\simeq 0.7\times 10^{-3}$. Similar to this case, 
we can always find cancellation regions in the cases of BP$b$ and BP$c$ as well, and thus 
conclude that $\rho_{bb}$-EWBG scenario is still consistent with the ACME18 bound. 
Note that here we set $c_\gamma=0.1$ while finding the constraints from ACME18 to illustrate 
$\irbb\sim0.15$ is still allowed for $\rtt\sim0.5$ for all the three BPs.
In the rest of the paper, however, we ignore the $c_\gamma$ dependence since it is insensitive to our collider study.

Here, we briefly discuss the EDMs of neutron and Mercury. 
Their current experimental values are respectively given by~\cite{Abel:2020gbr,Graner:2016ses}
\begin{align}
|d_n| & < 1.8 \times 10^{-26}~e~\text{cm}~(90\%~\text{C.L.}), \\
|d_{\text{Hg}}| & < 7.4\times 10^{-30}~e~\text{cm}~(95\%~\text{C.L.}).
\end{align}
On the theoretical side, the neutron EDM based on QCD sum rules is estimated as~\cite{Hisano:2015rna},
\begin{align}
d_n = -0.20d_u+0.78d_d+e(0.29d_u^C+0.59d_d^C)/g_3,
\end{align}
where $g_3$ is the $\text{SU}(3)_C$ gauge coupling and $d_q^C$ are the quark chromo EDMs defined by the 
operator $\mathcal{L}_{\text{CEDM}} = -(i/2)d_q^CG^{\mu\nu}\bar{q}\sigma_{\mu\nu}\gamma_5q$ with $G^{\mu\nu}$ 
representing $\text{SU}(3)_C$ field strength tensor.
We note that even though the cancellation mechanism can work in $d_n$ as well, it does not occur at the 
cancellation point of $d_e$. Using the same input parameters as in Fig.~\ref{fig:dThO_r} with $r=0.7\times10^{-3}$, 
we obtain $|d_n| = 2.4\times 10^{-29}~e~\text{cm}$, which is nearly 3 orders of magnitude below the current bound.
For the mercury EDM, on the other hand, we estimate it using formulas in Refs.~\cite{Ellis:2008zy,Ellis:2011hp,Cheung:2014oaa} 
assuming $d_{\text{Hg}}^{\text{I}}$ defined in Ref.~\cite{Ellis:2011hp} and find that $|d_{\text{Hg}}|=8.0\times 10^{-31}~e~\text{cm}$, 
which is smaller than the current bound by about 1 order of magnitude.

We note in passing that a future measurement of the proton EDM could be another good prober of $\rho_{bb}$-EWBG. The experimental 
sensitivity of the proton EDM at IBS-CAPP~\cite{Haciomeroglu:2018nre} and BNL~\cite{Anastassopoulos:2015ura} is $|d_p|\sim 10^{-29}~e~\text{cm}$.
As is the case of neutron EDM discussed above, the proton EDM can be estimated by use of the QCD sum rules as~\cite{Hisano:2015rna}
\begin{align}
d_p = 0.78d_u-0.20d_d+e(-1.2d_u^C-0.15d_d^C)/g_3.
\end{align}
With this, it is found that $|d_p|=6.1\times 10^{-29}~e~\text{cm}$ for the parameters used in Fig.~\ref{fig:dThO_r} with $r=0.7\times10^{-3}$.
Therefore, the future measurement of $d_p$ could access the $\rho_{bb}$-EWBG parameter space regardless of the $d_e$ cancellation.

\subsection{Direct search limits}
There exist several direct search limits from ATLAS and CMS that may restrict the 
parameter space of $\rbb$, even for $c_\gamma=0$ and $\rtt=0$.
The coupling $\rbb$ receives several constraints from heavy Higgs boson searches at the LHC. 
In particular, Refs.~\cite{Sirunyan:2018taj,ATLAS:2019jzx,Aaboud:2018cwk,Sirunyan:2019arl,Sirunyan:2020hwv}
are relevant to our study. We find that the most stringent
constraint arises from CMS search  
involving heavy Higgs boson production in association with at least one $b$-jet 
and decaying into $b\bar b$ pair based on 13 TeV 35.7 \fbi data~\cite{Sirunyan:2018taj}. 
The CMS search provides a model independent 95\% CL upper limits on the 
$\sigma(pp\to b A/H +X)\cdot\mathcal{B}(A/H\to b \bar b)$ in the mass range beginning from 
300 GeV to 1300 GeV. We first extract~\cite{extrac} corresponding 95\% CL upper limit
$\sigma(pp\to b A/H +X)\cdot\mathcal{B}(A/H\to b \bar b)$ 
for our three BPs. Taking a reference $|\rbb|$ value, we then estimate 
the production cross sections of $pp\to b A/H +X $ at the leading order (LO) 
utilizing Monte Carlo event generator MadGraph5\_aMC@NLO~\cite{Alwall:2014hca} (denoted as MadGraph5\_aMC) with
the default parton distribution function (PDF) NN23LO1 set~\cite{Ball:2013hta} for the BPs.
As the analysis does not veto additional activity in the event~\cite{Sirunyan:2018taj}, we therefore 
include contributions from $gg \to b \bar b A/H$ along with
$bg\to b A/H$ while estimating the cross sections. These cross sections 
are finally rescaled by $|\rbb|^2\times\mathcal{B}(A/H\to b \bar b)$, assuming $\mathcal{B}(A/H\to b \bar b)=100\%$,
to obtain the corresponding 95\% CL upper limits on $|\rbb|$. It is found
that $|\rbb|\gtrsim 0.6$ is excluded for BP$a$ at 95\% CL and likewise,
the regions where $|\rbb|\gtrsim 0.7$ are ruled out for both BP$b$ and BP$c$. 
These upper limits are rather weak and would be further weakened by $\mathcal{B}(A\to Z h)$ and $\mathcal{B}(A/H\to t \bar t)$.
The limits are even weaker from a similar search performed by ATLAS~\cite{ATLAS:2019jzx}. 
We note that while estimating the 
upper limit on $\rbb$ we set all $\rho_{ij}=0$ for simplicity. In general, we remark that nonzero $\rho_{ij}$
would further alleviate these upper limits.
Further, $\rbb$ coupling can induce $pp\to t(b)H^\pm$ process which is proportional to $V_{tb}$ (see Eq.~\eqref{eff}).
These processes are extensively searched by ATLAS~\cite{Aaboud:2018cwk} and 
CMS~\cite{Sirunyan:2019arl,Sirunyan:2020hwv}
with $H^+/H^- \to t \bar b/ \bar t b$ decays.
We find that the constraints are weaker for all the three BPs, however, 
as we see below these searches would provide sensitive probe to $\rtt$. The effective model is implemented in the
FeynRules~2.0~\cite{Alloul:2013bka} framework.

We now turn to constraints on $\rtt$.
As $\rtt$ can also induce $V_{tb}$, the searches $pp\to\bar t (b) H^+$ followed
by $H^+\to t \bar b$~\cite{Aaboud:2018cwk,Sirunyan:2019arl,Sirunyan:2020hwv}
would also be relevant. The ATLAS search~\cite{Aaboud:2018cwk} is based 
on 36 fb$^{-1}$ $\sqrt{s}=13$ TeV dataset, which provides model independent 95\% CL upper limit on
$\sigma(pp\to \bar t b H^+)\times  \mathcal{B}(H^+  \to t \bar b)$ 
from $m_{H^\pm}=200$ GeV and  2 TeV. Similar searches are also performed by CMS
based on $\sqrt{s}=13$ TeV 35.9 fb$^{-1}$ dataset~\cite{Sirunyan:2019arl,Sirunyan:2020hwv}. 
These searches provide 95\% CL 
upper limit on $\sigma(pp\to \bar t H^+)\times  \mathcal{B}(H^+  \to t \bar b)$
for $m_{H^\pm}=200$ GeV and 3 TeV in leptonic~\cite{Sirunyan:2019arl} and, 
combining leptonic and all-hadronic final states~\cite{Sirunyan:2020hwv}. 
Like before, the nonvanishing $\rho_{tt}$ enhanced by $V_{tb}$ 
can induce such process, leading to stringent constraints. To find the constraints, 
as done before, we calculate the cross sections $\sigma(pp\to \bar t b H^+)\times (H^+  \to t \bar b)$ 
at LO for a reference $|\rho_{tt}|$ for the three BPs
via MadGraph5\_aMC. These cross sections are then 
rescaled by $|\rtt|^2\times\mathcal{B}(H^+ \to t \bar b)$ to get the corresponding 95\% CL upper limits on $|\rtt|$. 
The extracted~\cite{extrac} 95\% CL upper limits from ATLAS search~\cite{Aaboud:2018cwk} on 
$\rtt$ for the three BPs are $|\rtt|\gtrsim 0.7,~0.8$ and 1, respectively, while the limits from CMS~\cite{Sirunyan:2020hwv} are 
much stronger, which read as   $|\rtt|\gtrsim 0.6,~0.61$ and 0.61, respectively.
We remark that the constraints from CMS search with leptonic final state~\cite{Sirunyan:2019arl}
is mildly weaker than the search with combined leptonic and all-hadronic final states~\cite{Sirunyan:2020hwv}.
We also note that all the $\rho_{ij}$ except for $\rtt$ are assumed to be zero when extracting the upper limits for the sake of simplicity.
Therefore if other $\rho_{ij}$ are turned on, the limits on $\rtt$ in general becomes weaker due to dilution 
from other branching ratios of $H^\pm$.

The ATLAS~\cite{Aaboud:2017hnm} and CMS~\cite{Sirunyan:2019wph} search for heavy Higgs 
via $gg\to H/A \to t \bar t$ would also constrain $\rho_{tt}$. 
The ATLAS~\cite{Aaboud:2017hnm} result is based on 20.3 fb$^{-1}$ data at 8 TeV, which
provides exclusion limits on $\tan \beta$ vs $m_A~(\mbox{or,}~m_H)$ in type-II 2HDM framework 
starting from $m_A$ and $m_H=500$ GeV. The CMS search is based on 35.9 fb$^{-1}$ data at $\sqrt{s}= 13$ TeV,
which provides upper limit on coupling modifier (see Ref.~\cite{Sirunyan:2019wph} for definition)
the $m_A$ ($m_H$) from 400 GeV to 750 GeV based on different values of $\Gamma_A/m_A$ ($\Gamma_H/m_H$)
ratios. Given the values of $m_A$ and $m_H$ of the BPs, ATLAS search can
only constrain BP$c$ via $gg\to H\to t \bar t$.
Reinterpreting the ATLAS exclusion limit~\cite{Aaboud:2017hnm}, we find that
$|\rho_{tt}|\gtrsim 0.8$ is excluded for BP$c$ at 95\% CL.
On the other hand, for CMS search~\cite{Sirunyan:2019wph}, $gg\to A \to t \bar t$
can constrain $|\rtt|$ only for BP$c$, whereas $gg\to H \to t \bar t$ constrains
all three BPs. We find that CMS $gg\to A \to t \bar t$ search excludes the region of $|\rho_{tt}| \gtrsim 1~(1.1)$ at 95\% CL for BP$c$ 
if $\Gamma_A/m_A=5\%$ ($\Gamma_A/m_A=10\%$). The $gg\to H \to t \bar t$ search places the constraints that
$|\rtt|\gtrsim 1.6~(2.1)$, $1.2~(1.4)$ and $1.2~(1.3)$ if $\Gamma_H/m_H=5\%$ ($\Gamma_H/m_H=10\%$) at 95\% CL for the three BPs respectively.
We remark that these upper limits provided by both the collaborations
assume that $m_A$ and $m_H$ are decoupled from each other. 
Although $m_A$ and $m_H$ are separated sufficiently, this
is not the case for any of the BPs chosen, as can be seen from Table~\ref{bench}. Therefore,
the actual upper limits extracted here would be mildly stronger.

Moreover, $\rho_{tt}$ would also receive constraint from CMS search 
for SM four-top production~\cite{Sirunyan:2019wxt}. The search is performed with 13 TeV 137 \fbi dataset
and provides 95\% CL upper limits on $\sigma(pp \to t \bar t A/t \bar t H)\times\mathcal{B}(A/H \to t \bar t)$ 
for $350~\text{GeV}\le m_{A/H} \le 650~\text{GeV}$. The search also includes 
contributions from $\sigma(pp\to t W A/H, t q A/H)$ followed by $A/H\to t \bar t$, 
which can also be induced by $\rho_{tt}$. To understand how strong the constraints could be, 
we generate these cross sections at LO by MadGraph5\_aMC for a reference value of $|\rho_{tt}|$ 
setting all other $\rho_{ij} = 0$, and then rescale simply by $|\rho_{tt}|^2 \times\mathcal{B}(A/H \to t \bar t)$.
Again, given the masses of BPs, the $\sigma(pp \to t \bar t A)\times\mathcal{B}(A \to t \bar t)$ search
can constrain only BP$c$, for which $|\rtt|\gtrsim 0.8$ excluded at 95\% CL. However, 
$\sigma(pp \to t \bar t H)\times\mathcal{B}(H \to t \bar t)$ can constrain all three BPs,
and we find that the regions of $|\rtt|\gtrsim 0.9$, 0.8 and 1 are excluded at 95\% CL, respectively,
where $\mathcal{B}(A/H \to t \bar t)=100\%$ is assumed. However, it should be noted that the presence of $\mathcal{B}(A/H\to b \bar b)$ and $\mathcal{B}(A\to Zh)$ would alleviate the limits.
As in the case of $gg\to H/A \to t \bar t$, the search here also assumes that $m_A$ and $m_H$ are decoupled from each other.
Therefore, one expects the limits to be mildly stronger for all the three BPs. 

We finally conclude that for $\rrbb\sim0$,
$c_\gamma=0.1$, $|\irbb|\sim 0.15$ and $\rtt\sim0.5$ are well allowed by the current measurements
for the mass spectrum under consideration. We take these values as 
representative values for our analysis with $\rrbb=0$.\footnote{As discussed before,
both $\rbb$ and $\rtt$ can induce $pp\to t(b)H^\pm$ process, hence,
the constraints from Ref.~\cite{Sirunyan:2020hwv} would become stronger if both the couplings are nonvanishing. 
However, we have checked that $\rtt\sim0.5$ is allowed for $|\irbb|\sim 0.15$ for all the three BPs.}
Under the assumptions, i.e., setting all $\rho_{ij}=0$ except $\rbb$ and $\rtt$, the total decay width of $A$ can be 
nicely approximated as the sum of the partial widths of $A \to b \bar b$ and $A \to Zh$ for BP$a$ and BP$b$ and,
$A \to b \bar b$, $A \to Zh$ and $A \to t \bar t$ for BP$c$. 
The total decay widths of $A$ are 0.4 GeV, 0.5 GeV and 4.97 GeV 
respectively for the three BPs with $|\irbb|\sim 0.15$ and
$\rtt\sim0.5$. The corresponding branching ratios are presented
in Table~\ref{branch}. Note that when calculating decay
widths and branching ratios of $A$, we neglect tiny loop induced decays such as $A\to \gamma\gamma$, $A\to Z\gamma$ etc.

\begin{table}[t]
\centering
\begin{tabular}{|c| c| c |c |c  c }
\hline
  BP  &  \hspace{0.05cm} $\mathcal{B}(A \to b\bar b)$  \hspace{0.05cm} &  \hspace{.05cm}  $\mathcal{B}(A\to Z h)$ \hspace{.05cm} 
  & \hspace{0.05cm} $\mathcal{B}(A \to t\bar t)$  \hspace{0.05cm}\\
\hline
\hline
     $a$                      &    0.95            &   0.05              & ---  \\
     $b$                      &    0.89            &   0.11             & ---     \\
     $c$                      &    0.12            &   0.04             & 0.84 \\            
\hline
\end{tabular}
\caption{Branching ratios of $A$ for the benchmark points in Table~\ref{bench} with $\rrbb=0.0$, $|\irbb|=0.15$ and $|\rtt| = 0.5$.} 
\label{branch}
\end{table}

\section{Collider Signatures}\label{coll}
\subsection{The \bzh process}\label{bZh}

In this subsection we study the discovery potential of \bzh process at 14 TeV LHC.
The process can be searched via $pp\to b A +X \to b Zh +X$ followed by $Z\to \ell^+ \ell^-$ ($\ell = e,~\mu$)
and $h\to b \bar b$, comprising three $b$-jets, same flavor opposite sign lepton pair (denoted as the $bZh$ process).
There are several SM backgrounds for this final state topology.
The dominant backgrounds are $t\bar t+$jets, Drell-Yan+jets (DY+jets), $Wt+$jets, $t\bar tZ$+jets, $t\bar t h$, $tZ$+jets,
with subdominant contributions arise from four-top ($4t$), $t\bar t W$, $tWh$, $tWZ$ and $WZ$+jets. 
Backgrounds from $WW$+jets and $ZZ$+jets are negligibly small and hence not included. Note that one can also search 
for $h\to \gamma \gamma$ or $h\to \tau \tau$; however, we do not find them as promising.

The signal and background event samples are generated in $pp$ collision with $\sqrt{s}=14$ TeV CM energy 
at LO by MadGraph5\_aMC with NN23LO1 PDF set as done before and then 
interfaced with Pythia 6.4~\cite{Sjostrand:2006za} for hadronization and showering and finally fed 
into Delphes~3.4.2~\cite{deFavereau:2013fsa} for fast detector simulation adopting default ATLAS-based detector card.
We adopt MLM scheme~\cite{Mangano:2006rw,Alwall:2007fs} for matrix element and parton shower merging.  
Note that we have not included  backgrounds from  the fake and non-prompt sources in our analysis. 
Such backgrounds are not properly modeled in Monte Carlo simulations and requires data to estimate such contributions.

The LO $t\bar t+$jets and $Wt+$jets cross sections are normalized to 
NNLO (next-to-next-to LO) with NNLL (next-to-next-to leading logarithmic) corrections by factors
1.84~\cite{twiki} and $1.35$~\cite{Kidonakis:2010ux} respectively. We normalize the DY+jets background cross section 
to the NNLO QCD+NLO EW one by factor 1.27, which is obtained by utilizing FEWZ 3.1~\cite{Li:2012wna,Hou:2017ozb}.
The LO $t\bar t Z$,  $\bar tZ +$ jets, $t\bar t h$, $4t$ and  $t\bar{t} W^-$ ($t\bar{t} W^+$) cross sections are
adjusted to NLO ones by $K$-factors 1.56~\cite{Campbell:2013yla},
1.44~\cite{Alwall:2014hca}, 1.27~\cite{twikittbarh}, 2.04~\cite{Alwall:2014hca} and 1.35 (1.27)~\cite{Campbell:2012dh} respectively,
but $tWZ$ and $tWh$ both are kept at LO. Finally, the background $W^-Z+$jets is normalized to NNLO by factor 
2.07~\cite{Grazzini:2016swo}. For simplicity we assume the same 
QCD correction factors for the conjugate processes $tZj$ and $W^+Z+$jets,
while the signal cross sections are kept at LO.

\begin{figure*}[htb]
\centering
\includegraphics[width=.49 \textwidth]{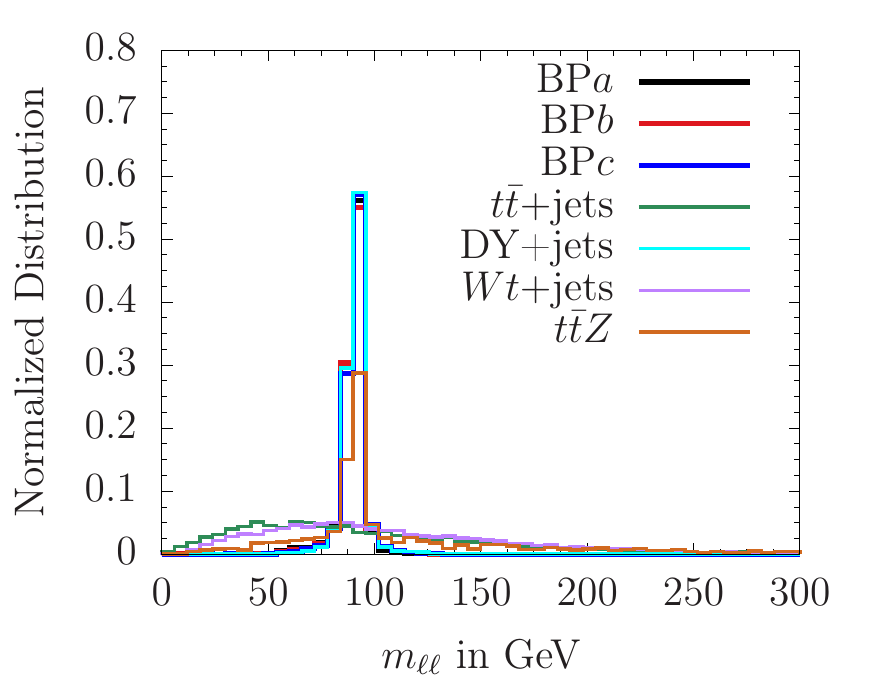}
\includegraphics[width=.49 \textwidth]{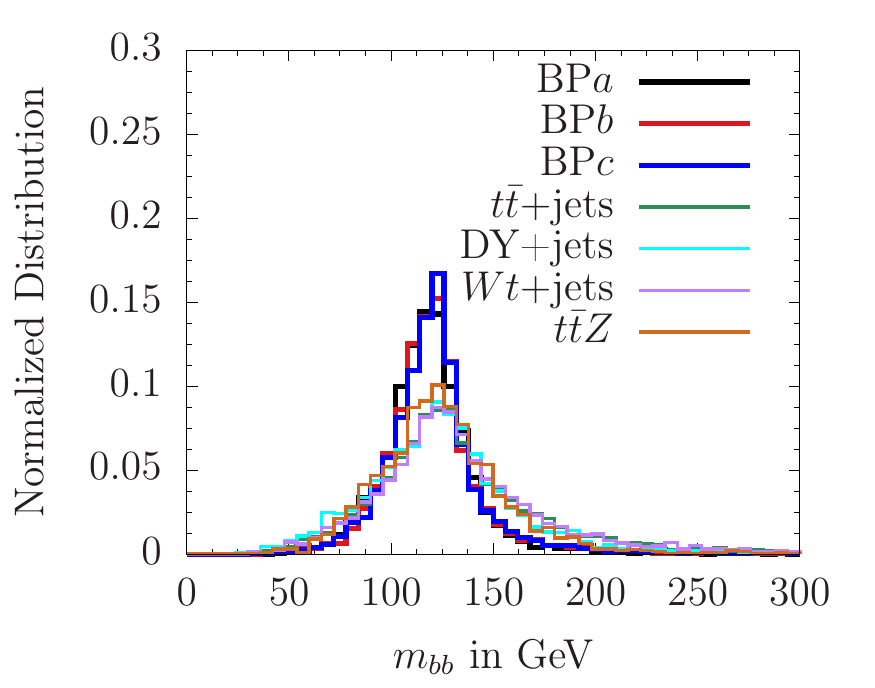}
\caption{The normalized $m_{\ell\ell}$ (left) and $m_{bb}$ (right) distributions 
of the three BPs and leading backgrounds for the $bZh$ process.}
\label{bzhdist1}
\end{figure*}
\begin{figure*}[htb]
\centering
\includegraphics[width=.5 \textwidth]{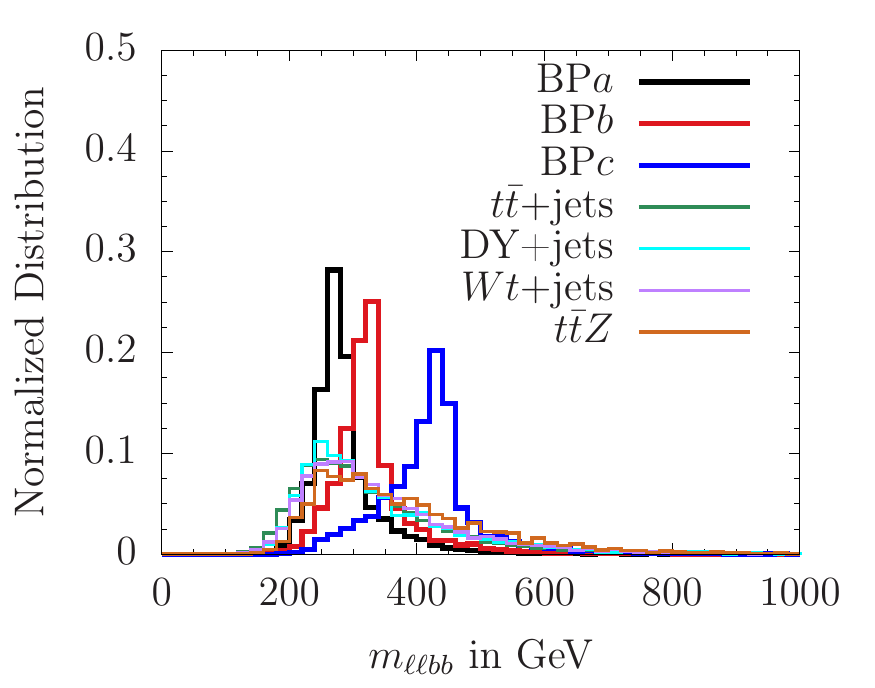}
\caption{The normalized $m_{\ell\ell bb}$ distributions 
of the signal and leading backgrounds for the $bZh$ process.}
\label{bzhdist2}
\end{figure*}

\begin{table*}[hbt!]
\centering
\begin{tabular}{|c |c| c| c| c | c| c| c |c |c|}
\hline
&&&&&&&&\\ 
      BP            & $t\bar t+$ jets  & $DY+$ jets   & $Wt+$ jets   &  $t\bar t Z$   &  $t\bar t h$    & $t Z+$ jets & Others & \ Total  \ Bkg.         \\
&&&&&&&& (fb)\\                                
\hline
\hline
&&&&&&&&\\
       $a$            &   0.178       & 0.533        & 0.226      & 0.023        & 0.008    & 0.007   & 0.003   & 0.978  \\ 
       $b$            &   0.087       & 0.272        & 0.084      & 0.018        & 0.003    & 0.006   & 0.001   & 0.471   \\  
       $c$            &   0.016       & 0.094        & 0.022      & 0.008        & 0.0002   & 0.002   &  0.0004 & 0.143  \\ 

\hline
\hline
\end{tabular}
\caption{The background cross sections (in fb) for the
the $bZh$ process after selection cuts at $\sqrt{s}=14$ TeV LHC.
Here we added together subdominant backgrounds $4t$, $t\bar t W$, $tWh$, $tWZ$ and $WZ$+jets as 
``Others''. The total background (Total Bkg.) yield is provided in last column.}
\label{bkgcompbzh}
\end{table*}
\begin{table}[hbt!]
\centering
\begin{tabular}{|c |c| c| c | c }
\hline 
                     BP          &  \ Signal \              &  \ Significance ($\mathcal{Z}$)     \\ 
                                 &       (fb)               &  300 (1000) fb$^{-1}$    \\                                      
\hline
\hline

                     $a$            &  0.17                    &  2.9 (5.3)   \\ 
                     $b$            &  0.228                   &  5.4 (9.8) \\
                     $c$            &  0.027                   &  1.2 (2.2)   \\ 
\hline
\hline
\end{tabular}
\caption{The signal cross sections after selection cuts and the corresponding 
significances of the $bZh$ process with 300 (1000) fb$^{-1}$ integrated luminosity.}
\label{signibzh}
\end{table}

To reduce backgrounds, we adopt following event selection criteria: 
each event should contain a same flavor opposite sign lepton pair and 
at least three $b$-tagged jets. The transverse momenta ($p_T$) of the leading and subleading leptons should be 
$> 28$ GeV and $> 25$ GeV respectively, whereas $p_T$ for all three $b$-jets should be $> 20$ GeV. 
The pseudo-rapidity ($|\eta|$) of the leptons and all three $b$-jets
are required to be $<2.5$. 
The jets are reconstructed by anti-$k_T$ algorithm with radius parameter $R = 0.4$.
The separation $\Delta R$ between any two $b$-jets, a $b$-jet and a lepton
and between two leptons should be $> 0.4$.
In order to reduce the $t\bar t$+jets background, we veto events having missing transverse energy ($E_T^{\rm{miss}}$) $> 35$ GeV.
The invariant mass of the two same flavor opposite charge leptons ($m_{\ell\ell}$) is needed to remain between $ 76 < m_{\ell\ell} < 100$ GeV, i.e., the $Z$ boson mass window. We then apply invariant mass for two $b$-jets $m_{bb}$ in a event.
As there are at least three $b$-jets in a event, more than one $m_{bb}$ combinations are possible;
the one closest to $m_h$ is selected and required to remain within $|m_h-m_{bb}|< 25$ GeV.
Further, we require the invariant mass $m_{\ell\ell bb}$ constructed from the two same flavor opposite charge leptons and
$b$-jets combination that passes the $m_{bb}$ selection to be within $|m_A- m_{\ell\ell bb}|< 50$ GeV. 
The normalized $m_{\ell\ell}$ and $m_{bb}$ distributions before application of any selection cuts are presented in Fig.~\ref{bzhdist1} while 
the same for $m_{\ell\ell bb}$ is shown in Fig.~\ref{bzhdist2}.
We adopt the $b$-tagging efficiency and $c$- and light-jets misidentification efficiencies of Delphes ATLAS based detector card.
The background cross sections after selection cuts of the three benchmark points are summarized in Table.~\ref{bkgcompbzh}, while the signal cross sections along with their corresponding
significances with the integrated luminosity $\mathcal{L}= 300$  and 1000 \fbi  are presented in Table~\ref{signibzh}.
The statistical significances are estimated using
$\mathcal{Z} = \sqrt{2[ (S+B)\ln( 1+S/B )-S ]}$~\cite{Cowan:2010js}, where
$S$ and $B$ are the numbers of the signal and background events.

Let us take a closer look at Table~\ref{signibzh}. We find very promising the discovery potential 
with sufficiently large $S/B$ ratio, especially for $m_A < 2m_t$.  
The achievable significance for the  BP$a$ and  BP$b$ are  $\sim 2.9\sigma$ ($\sim 5.3\sigma$) and
$\sim 5.4 \sigma$ ($\sim 9.8\sigma$) respectively with 300 (1000) fb$^{-1}$ integrated luminosity. The BP$c$ requires 
larger dataset due to fall in parton luminosity and suppression from $\mathcal{B}(A\to t \bar t)$ decay
and $\sim 2.2\sigma$ is possible with 1000 \fbi but could reach up to $\sim 3.8\sigma$ with the full high luminosity LHC (HL-LHC) dataset
(3000 \fbi integrated luminosity).

\subsection{The \bgba process}\label{btt}
We now discuss the discovery potential of $pp \to b A + X \to b t \bar t +X$, 
followed by semileptonic decay of at least one top quark, constituting 
three $b$-jets, at least one charged lepton ($e$ and $\mu$) and missing 
transverse energy ($E^{\rm{miss}}_T$) signature, which we denote as $3b1\ell$ signature.
Note that \bgba is only possible for BP$c$ as $m_A < 2 m_t$ but for BP$a$ and BP$b$
one can have $3b1\ell$ signature via $pp \to b H + X \to b t \bar t +X$. 
However, such signatures will be mild for the former two BPs due to suppression from $\mathcal{B}(H \to A Z)$. 
Further, $bg\to \bar t H^+ \to \bar t t \bar b$
process may also contribute to the same final state topologies, if at least one of the top decays semileptonically.
Such contribution could be moderate for all three BPs. In our analysis, however, we neglect them for simplicity.

There exist several SM backgrounds. The dominant backgrounds are 
$t\bar t+$jets, $t$- and $s$-channel single-top ($tj$), $Wt$, with subdominant backgrounds 
from $t\bar t h$ and, $t\bar t Z$ productions. Further, small contributions come from 
Drell-Yan+jets, $W+$jets, four-top ($4t$), $t\bar t W$, $tWh$, which are collectively denoted 
as ``Others''. We do not include backgrounds originating from non-prompt and fake sources. 
These backgrounds are not properly modeled in Monte Carlo event generators and one requires data to estimate
such contributions. 

Here we follow the same event generation procedure for signal and backgrounds as in previous subsections, i.e.,
via MadGraph5\_aMC followed by hadronization and showering in Pythia and with Delphes ATLAS based detector card for fast detector simulation.
The LO $t\bar t +$jets background cross section is normalized up to the
NNLO by a factor of $1.84$ while
$t$- and $s$-channel single-top cross sections are normalized by factors of 1.2 and 1.47, respectively~\cite{twikisingtop}. 
The LO $Wt+$jets background is normalized to the NLO cross section by a factor of 1.35, 
whereas the subdominant $t\bar t h$ and $t\bar t Z$ are corrected to corresponding NLO 
ones by factors of 1.27 and 1.56 respectively.
The DY+jets background is normalized to  NNLO cross sections by factor of 1.27.
Finally, the LO cross sections $4t$ and $t\bar t W$ are adjusted to the NLO ones by factors of 
2.04 and 1.35, respectively.
The $tWh$ and $W+$jets background are kept at LO.
For simplicity we assume the correction factors for the charge conjugate processes to be the same. 
We remark that the signal cross sections for all the three BPs are kept at LO.

\begin{figure*}[b]
\centering
\includegraphics[width=.45 \textwidth]{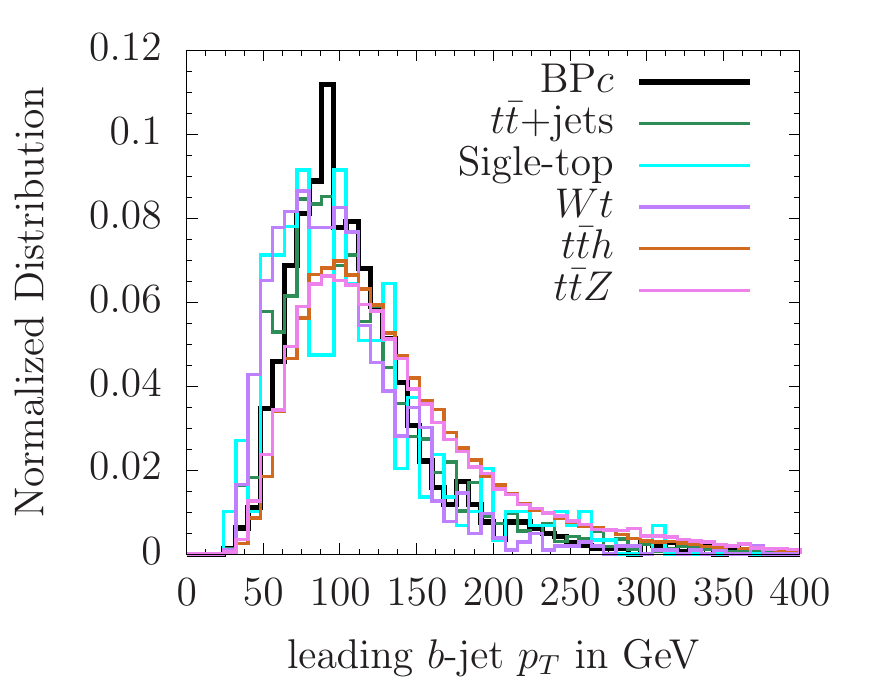}
\includegraphics[width=.45 \textwidth]{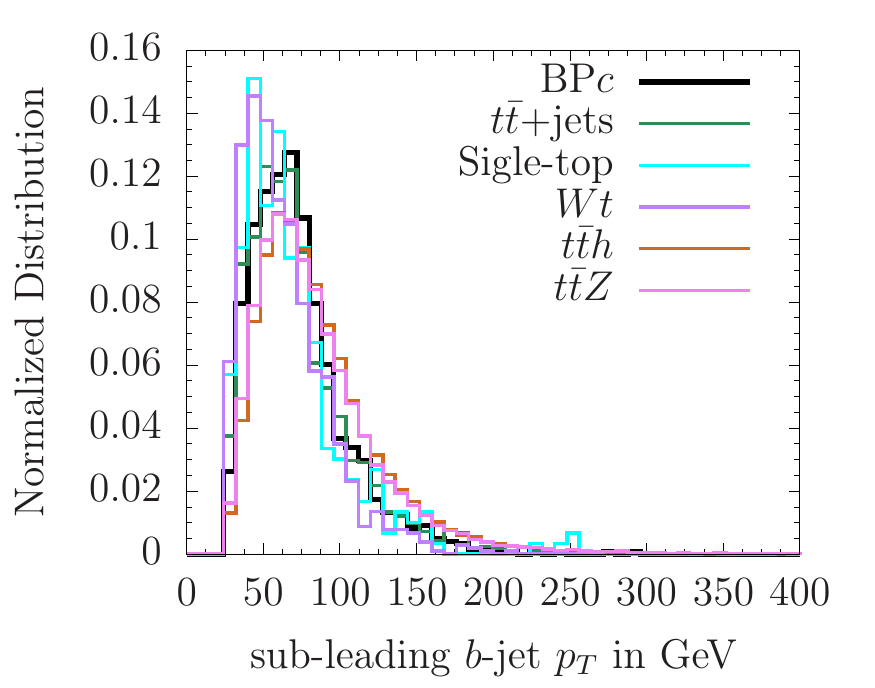}
\caption{The normalized $p_T$ distributions of the leading and subleading $b$-jets for the signal (BP$c$) and the leading backgrounds.}
\label{pt1}
\end{figure*}

\begin{figure*}[htbp!]
\centering
\includegraphics[width=.45 \textwidth]{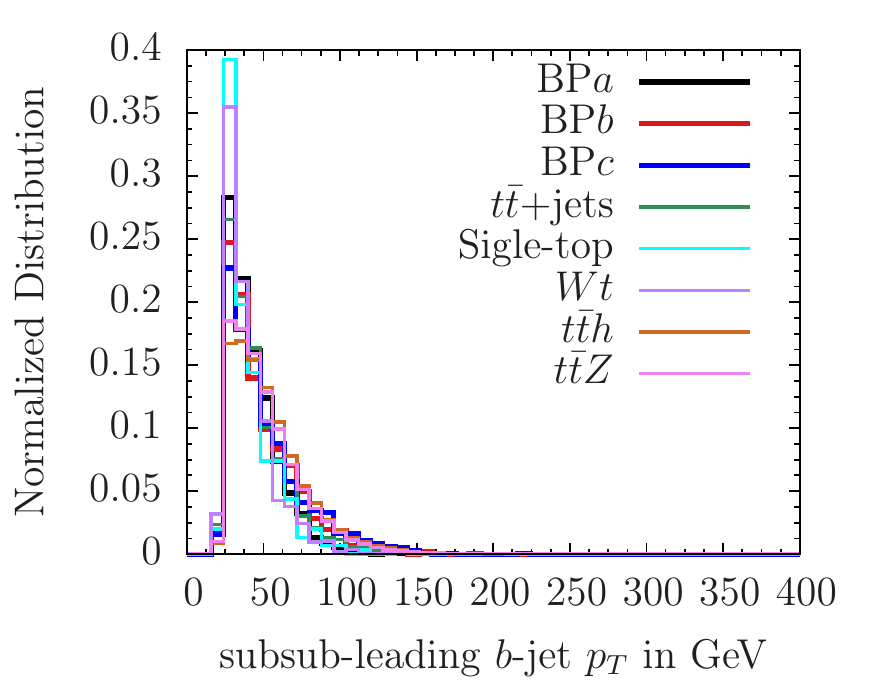}
\includegraphics[width=.45 \textwidth]{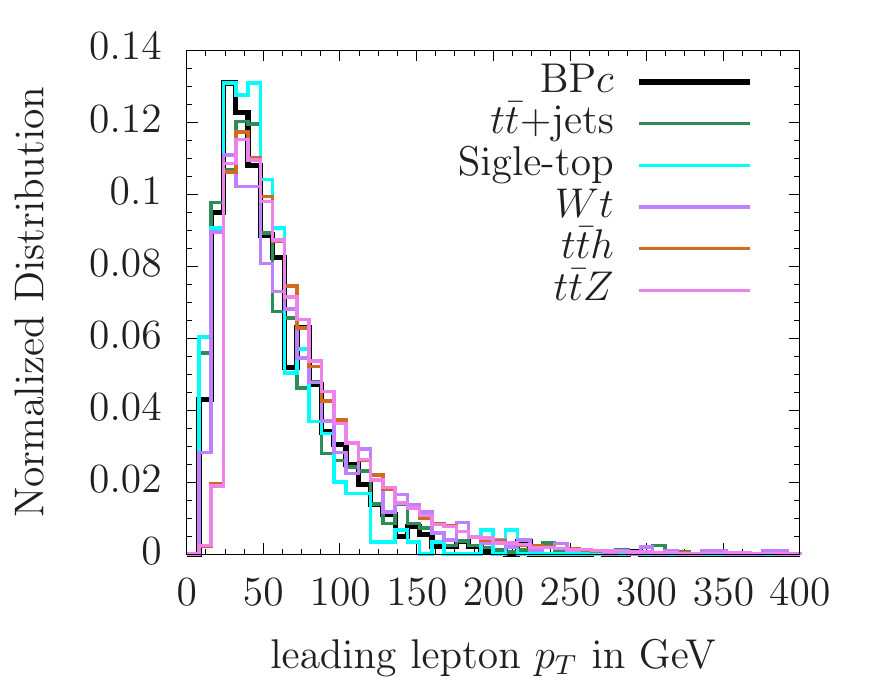}
\caption{The normalized $p_T$ distributions of the subsubleading $b$-jet (right) and leading lepton (left) for the signal (BP$c$) and leading 
backgrounds.}
\label{pt2}
\end{figure*}

\begin{figure*}[htbp!]
\centering
\includegraphics[width=.45 \textwidth]{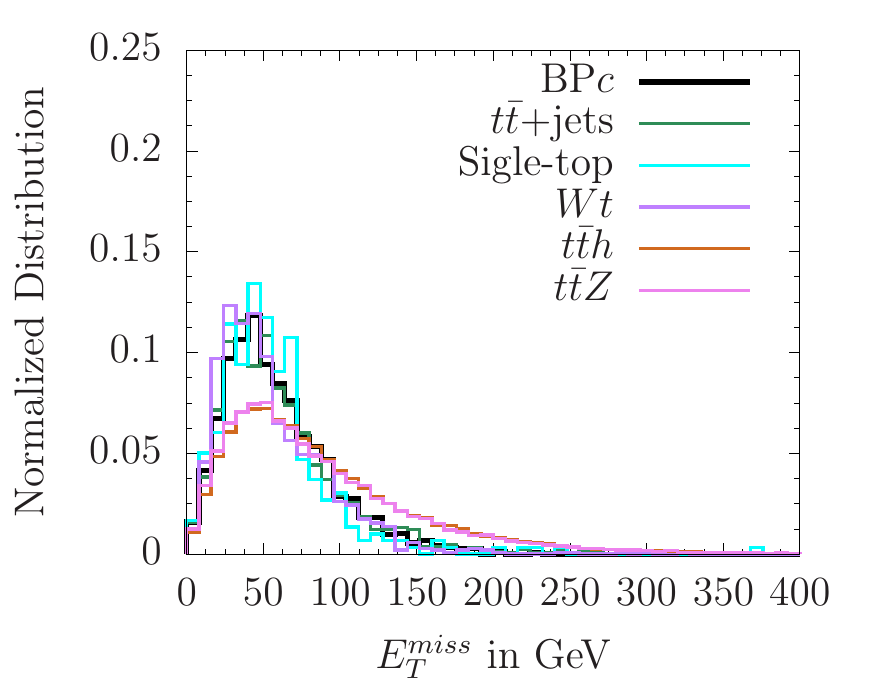}
\includegraphics[width=.45 \textwidth]{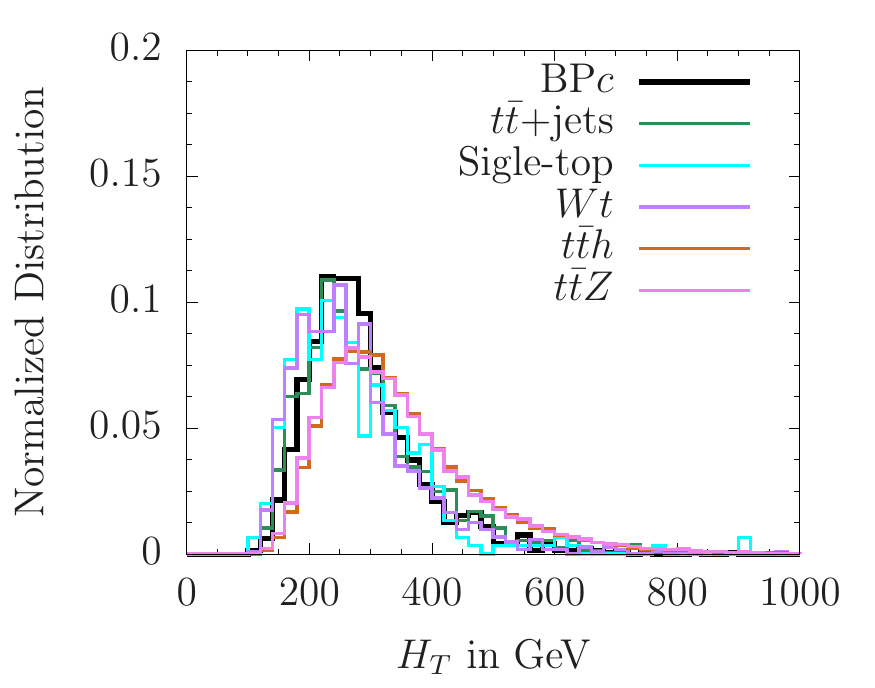}
\caption{The normalized missing energy $E^{\rm{miss}}_T$ (left) and $H_T$ (right) for the signal (BP$c$) and leading backgrounds.}
\label{ethtmiss}
\end{figure*}

The events are selected in a way such that they should contain at least 
one charged lepton ($e$ and $\mu$), at least three $b$-tagged and 
some $E^{\rm{miss}}_T$. The normalized transverse momentum ($p_T$) distributions 
of the leading and subleading $b$-jets for the signal and leading backgrounds are presented in Fig.~\ref{pt1}. 
The $p_T$ distributions for the subsubleading $b$-jet and 
leading lepton are plotted in Fig.~\ref{pt2}, while the normalized $E^{\rm{miss}}_T$ and $H_T$ (i.e., the scalar sum 
of $p_T$ of the leading charged lepton and the three leading $b$-jets)
distributions are shown in Fig.~\ref{ethtmiss}.
To reduce backgrounds we apply the following event selection cuts.
$p_T$ of all three $b$-jets should be $ p_T>20$ GeV, whereas that of the leading lepton should be $p_T> 25$ GeV. 
The absolute value of pseudo-rapidity ($|\eta|$)
of all three $b$-jets and lepton should be less than $2.5$. 
The minimum separation ($\Delta R$) between the lepton and any $b$-jet 
as well as that between any two $b$-jets are required to be 
greater than $0.4$. The $E^{\rm{miss}}_T$ in each event should be larger than $35$ GeV.
Note that in our exploratory study we do not optimize the selection 
cuts such as $p_T$, $\eta$, $E^{\rm{miss}}_T$ and $H_T$ for simplicity.
The signal and total background cross sections along with different components 
after the selection cuts are shown in Table~\ref{sigbkgcompba0}.

\begin{table}[hbt!]
\centering
\begin{tabular}{|c |c| c| c| c | c| c| c |c |c|}
\hline
&&&&&&&&\\ 
        BP   &  Signal     &   $t\bar t+$  jets & Single-top      & $Wt+$ jets     &  $t\bar t h$   &  $t\bar t Z$    & Others   & \ Total Bkg.  \  \\
&&&&&&&& (fb)\\                                
\hline
\hline
&&&&&&&&\\
        BP$c$         &   9.27       &  3953.49 & 98.93 &  77.93 & 10.72 &  4.13 & 30.85 & 4176.05 \\

\hline
\hline
\end{tabular}
\caption{The cross sections (in units of fb) of the signal (BP$c$) and different background 
components for $3b1\ell$ process after selection cuts at $\sqrt{s}=14$ TeV.}
\label{sigbkgcompba0}
\end{table}

The corresponding significance $\mathcal{Z}$ for the integrated luminosities 300 (3000) \fbi for the BP$c$ 
is $\sim 3.5\sigma$ ($\sim 7.9\sigma$). For the BP$a$ and BP$b$ from $pp \to b H + X \to b t \bar t +X$
are  $\sim 2.4\sigma$ and $\sim 2.8\sigma$ respectively for 3000 \fbi. Note that here we have not considered the systematic uncertainties
associated with the backgrounds, which could be considerable in particular for the largest $t\bar t+$ jets backgrounds.
As $S \ll B$ for the $3b1\ell$ process, in the presence of systematic uncertainties the $\mathcal{Z}$ formula 
simply becomes $\mathcal{Z} \approx S/\sqrt{B+\sigma_B^2}$. The $\sigma_B$ denotes the systematic uncertainty
that depends on the factor between control sample and the background in the signal region. The value of $\sigma_B$
is very much analysis~\cite{efejack} dependent. Here we assume that the systematic uncertainty
arise only from $t\bar t+$ jets backgrounds for simplicity and neglect the same for other subdominant backgrounds and take 
two different for $\sigma_B$ for illustration. E.g., if $\sigma_B = \sqrt{B}$
the significance of BP$c$ reduces to $\sim 5.5\sigma$ whereas for $\sigma_B=0.1\;B$ the significances goes below $1\sigma$
with HL-LHC dataset. Similarly, for $\sigma_B = \sqrt{B}$ the significances of BP$a$ and BP$b$ from $pp \to b H + X \to b t \bar t +X$ 
reduces to $\sim 1.7\sigma$ and $\sim 2\sigma$ respectively but much below $1\sigma$ if $\sigma_B = 0.1\;B$.
Therefore we remark that the $3b1\ell$ process is promising, but one needs precise understanding of the background
systematics. 

\subsection{The \gga process}\label{ttA}
We now briefly discuss the discovery potential of $gg\to t \bar t A\to t \bar t b \bar b$ process.
The process can in principle probe the parameter space for $\rho_{bb}$-EWBG mechanism. 
We search this process via $pp\to t \bar t A +X\to t \bar t b \bar b +X$, followed by 
at least one top quark decaying semileptonically  i.e., with four $b$-jets, 
at least one charged lepton and $E_T^{\rm{miss}}$ signature. The final state topology receives mild 
contributions from inclusive $pp\to W t A +X$ and $pp\to t j A +X$ processes. 
As we show below, the signature is not promising as opposed to \bgba process.

We generate events at LO as in $3b1\ell$ process, i.e., 
via MadGraph5\_aMC followed by hadronization and showering in Pythia and finally incorporate
the detector effects of Delphes ATLAS based detector card. The dominant backgrounds arise from 
the $t\bar t+$  jets, Single-top and $Wt+$ jets, whereas $t\bar t h$, $t\bar t Z$ and $4t$ constitute
subdominant backgrounds. We assume the same QCD corrections factor as in $3b1\ell$ process for simplicity. 

\begin{table}[hbt!]
\centering
\begin{tabular}{|c |c| c| c| c | c| c| c |c |c|}
\hline
&&&&&&\\ 
             BP     & Signal &  $t\bar t+$  jets & Single-top      & $Wt+$ jets   & Others   & \ Total Bkg.  (fb) \          \\
&&&&&&\\                                
\hline
\hline
&&&&&&\\
             BP$a$  &  0.2           &  229.2  & 4.5 &  2.8 & 7.8 & 244.3    \\
             BP$b$  &  0.14          &  204.9  & 4.0 &  2.3 & 8.2 & 219.4   \\
             BP$c$  &  0.01          & 157.8   & 2.7 &  1.9 & 5.3 & 167.7   \\
\hline
\hline
\end{tabular}
\caption{The cross sections (in units of fb) of three BPs and the different background components for the $4b1\ell$
process after selection cuts at $\sqrt{s}=14$ TeV.}
\label{bkg}
\end{table}

To reduce the background, we use the following event selection cuts. The events are selected so 
that they contain at least one lepton ($e$ and $\mu$), at least four jets with at 
least four are $b$-tagged and some missing $E_T^{\rm{miss}}$ 
(denoted as $4b1\ell$ process). The lepton is required to have $p_T > 25$ GeV and $|\eta| < 2.5$. 
For any jet in the event $p_T> 20$ GeV and $|\eta| < 2.5$. $E_T^{\rm{miss}}$ in each event is required to be greater than $35$ GeV.
The separation $\Delta R$ between any two jets as well as that between a jet and a lepton should be larger than $0.4$. Finally, 
we construct all possible combinations of the invariant mass $m_{jj}$ from the  four leading jets and demand that 
the one closest to $m_A$ should lie between $|m_A-m_{jj}|< 50$ GeV. The impact of these cuts on the signal 
and background processes are summarized in Table~\ref{bkg}.

We find that the achievable significance for all three BPs of the $gg\to t \bar t A\to t \bar t b \bar b$ process
are below $\sim 1 \sigma$ with 3000 \fbi integrated luminosity, which is rather low.
This means that no meaningful constraints can be extracted unless both ATLAS and 
CMS data are added. It should be remarked that since we use the same QCD
correction factors for the backgrounds as in \bgba process, there are greater 
uncertainties in these cross sections.  

Before closing this section, we discuss the impact of the other $\rho_{ij}$ couplings. So far,
we have set all $\rho_{ij} =0$ except $\rho_{bb}$ and $\rho_{tt}$. 
Presence of the other $\rho_{ij}$ couplings open up other decay modes of $A$,
which in principle may reduce the achievable significances summarized in 
previous subsections. For instance, if $\rho_{\tau\tau}\sim \lambda_\tau$, 
it would induce $A \to \tau^+ \tau^-$ decay. However, the significances remain practically same for all the three BPs.
Moreover, $\rho_{tc}\sim 0.3\text{-}0.4$ is still allowed by current data and would induce $cg\to t A/ tH \to t t \bar c$ (same-sign top)~\cite{Kohda:2017fkn,Hou:2018zmg} 
(see also Refs.~\cite{Hou:1997pm,Iguro:2017ysu,Altmannshofer:2016zrn}) 
and $cg\to t A/ tH \to t t \bar c$ (triple-top) signature, which might emerge in the Run-3 of LHC.

\section{Discussion and Summary}\label{summ}
We have analyzed the available constraints and prospect of probing  EWBG driven by the extra bottom Yukawa coupling $\rbb$
at the ongoing and future experiments. The parameter space receives
meaningful constraints from $h$ boson coupling measurements, $\bsg$, $\dcp$ of \ebsg, electron EDM measurement
and heavy Higgs searches at the LHC. We primarily focused on sub-TeV $m_A$, $m_H$ and $m_{H^\pm}$ with mixing 
angle $c_\gamma \sim 0.1$, which is required by EWBG~\cite{ewbg_2hdm,ewbg_susy,ewbg_singletSM,Modak:2018csw}. 
The constraints would be improved, e.g., on $c_\gamma$, $\rbb$ and $\rtt$ at the HL-LHC~\cite{Modak:2018csw,Hou:2018uvr} if no discovery is made.
This would allow us to probe even larger part of the parameter space of $\rho_{bb}$-EWBG.

Taking three benchmark points for illustration, two below $2m_t$ threshold and one above,
we have shown that a discovery is possible at the LHC via $\rbb$ induced \bzh process for 
$m_A < 2 m_t$. We find that the process may emerge in the Run 3 of LHC if $250~\mbox{GeV}\lesssim m_A \lesssim 350$ GeV. 
With a simple rescaling of the significances in Table~\ref{signibzh}, we find that $|\irbb|\gtrsim 0.05$ and $\gtrsim 0.04$ 
can be excluded for BP$a$ and BP$b$ with full HL-LHC dataset. Those are below the nominal value $|\irbb|\gtrsim 0.058$ required 
EWBG. For $m_A > 2 m_t$, a discovery may happen via \bgba. Note that one may also have the $bg\to b H \to b h h$ process,
which could be sensitive if $H$ is lighter than $A$ and $H^\pm$. The process is being studied elsewhere.
A discovery of the \bgba process is possible via $3b1\ell$ signature if $m_A\sim 450$ GeV and $\rtt\sim 0.5$ but
requires controlling of the systematics of $t\bar t$-jets background.
Additionally, we have also investigated the potential for 
\gga process but find it below the sensitivity even at the HL-LHC.

In principle the $pp \to b A + X$ process can also be 
induced by $\rho_{bd}$, $\rho_{db}$, $\rho_{bs}$ and $\rho_{sb}$ at the LHC. 
However, due to severe constraints arising from $B_d$ and $B_s$ mixings~\cite{Chen:2018hqy} their
impacts are typically inconsequential. In addition if the charm quark gets misidentified as $b$-jet, a sizable $\rho_{cc}$ 
can also mimic similar signature in $pp$ collision via $cg\to c A \to c t \bar t$ process. 
We remark that such possibilities can be disentangled by the simultaneous application of $b$- and $c$-tagging 
on the final state topologies~\cite{Hou:2018npi}.

Although the discovery is possible at the LHC,
to attribute it to $\rho_{bb}$-EWBG mechanism is beyond the scope of LHC as 
information of the CP-violating phase of $\rho_{bb}$ is lost in $pp$ collision. 
In this regard, $\dcp$ of $\bsg$ would provide very sensitive probe for the $\irbb$ even though 
the observable has uncertainties associated with the hadronic parameter $\tilde{\Lambda}_{78}$. 
While finding the constraints in Fig.~\ref{scanim}, we assumed $\tilde{\Lambda}_{78}=89$ MeV, 
which is the average of $17~\text{MeV}\le \tilde{\Lambda}_{78}\le190~\text{MeV}$~\cite{Benzke:2010tq}. 
However, if $\tilde{\Lambda}_{78}$ is taken as its upper range, the constraint becomes 
much severe for $\irbb$. Furthermore, on the experimental side,
projected Belle II accuracy of $\dcp$ measurement is $\sim5\%$~\cite{Kou:2018nap}. 
Therefore, more precise estimation of $\tilde{\Lambda}_{78}$ together with Belle II measurement
can stringently probe the parameter space of $\irbb$ unless the CP-violating
phases of $\rho_{tt}$ and $\rho_{bb}$ are aligned~\cite{Fuyuto:2019svr} in which $\dcp=0$. 
In such a case, measurements of EDMs play a pivotal role in probing $\irbb$.

The unprecedented electron EDM constraint set by ACME Collaboration in
2018 reduces most EWBG scenarios to despair. We updated our previous
analysis done in Ref.~\cite{Modak:2018csw} including all the relevant Barr-Zee diagrams. 
Because of the significant contributions arising from the diagrams involving $\rho_{ee}$,
the cancellation mechanism can be effective. It was found that the electron EDM 
cancellation in $\rho_{bb}$-EWBG belongs to the unstructured cancellation category
in which the diagonal hierarchical structures of $\rho^F_{ij}$ are much different
from those of the SM Yukawa couplings, which is in stark contrast to the case 
in $\rho_{tt}$-EWBG that can accommodate the structured cancellation~\cite{Fuyuto:2019svr}.
Nonetheless, the viable parameter space of $\rho_{bb}$-EWBG still exists.
Besides the extreme fine tuning of the parameters, $\rho_{bb}$-EWBG 
would be confirmed or ruled out if the electron EDM is improved down to $\sim10^{-30}~e~\text{cm}$ level. 
Moreover, as discussed in Sec.~\ref{sec:edms}, the future measurement of 
the proton EDM could play a complementary role in probing $\rho_{bb}$-EWBG.

So far we have not discussed the uncertainties.
As a first estimate, uncertainties arising from factorization scale ($\mu_F$) and
renormalization scale ($\mu_R$) dependences are not included in our LO cross section estimations.
In general, the LO $bg\to bA/bH$ processes have $\sim25-30\%$ scale uncertainties for $m_{A/H}\sim (300-400)$ GeV 
if bottom quark with $p_T > (15-30)$ GeV and, $|\eta|<2.5$~\cite{Campbell:2002zm} 
(see also~\cite{Dicus:1998hs,Maltoni:2005wd,Harlander:2003ai}). 
It has been found that~\cite{Maltoni:2003pn} the LO cross sections calculated with LO PDF set
CTEQ6L1~\cite{Pumplin:2002vw} have relatively higher factorization scale dependence.
Therefore, we remark that the LO cross sections in our analysis, which we
estimated with LO NN23LO1 PDF set, might have same level of uncertainties.
It has also been found that~\cite{Maltoni:2003pn} for $\mu_F\approx m_{A}~(\mbox{or}~m_H)$  
the corrections to the LO cross sections could be large negative ($\sim-70\%$), whereas
for the choice of $\mu_F\approx m_A/4 ~(\mbox{or}~m_H/4)$ the corrections are mild; which indicates that the 
$\mu_F\approx m_A/4$ is the relevant factorization scale. Furthermore, the cross section uncertainties from  factorization scale and 
renormalization  choices are found to be particularly small at $\mu_R=m_A$ and if varied from
$\mu_R=m_A/2$ to $\mu_R=2m_A$, along with $\mu_F = m_A/4$ and 
varied from $\mu_F=m_A/8$ to $\mu_F=m_A/2$~\cite{Maltoni:2003pn}.
In addition, our analysis does not include PDF uncertainties, 
which could be in general significant for any 
bottom-quark initiated process as discussed, e.g., in Ref.~\cite{Maltoni:2012pa}. 
Detailed discussions on different PDFs and associated uncertainties for the LHC are
summarized in Ref.~\cite{Butterworth:2015oua}. These typically would induce some
uncertainties in our results. We leave out the detailed estimation of these uncertainties for future work.

In summary, we have explored the possibility of electroweak baryogenesis 
induced by the extra bottom Yukawa coupling $\rbb$ 
via direct and indirect signatures at the collider experiments. We find that the discovery is possible at the high luminosity LHC if $250~\mbox{GeV}\lesssim m_A \lesssim 350$ GeV.
We also find that heavier mass ranges can also be probed via $bg\to bA \to b t \bar t$ but the process is associated with larger
uncertainties. While LHC can indeed discover the process, however, the 
information of the CP-violating phase of $\rbb$ can only be probed via $\dcp$ of $\bsg$ or the 
EDM measurements of the electron, neutron and mercury though the latter two have the less probing power to date. 
For completeness we also studied $gg\to t \bar t A\to t \bar t b \bar b$ process and found that it is not promising. 
In conclusion, together with the electron EDM measurement and/or $\dcp$ of $\bsg$ decay, the discovery 
of \bgba process may help us to understand the mechanism behind the observed matter-antimatter asymmetry of the Universe.

\vskip0.2cm
\noindent{\bf Acknowledgments.--} \
TM thanks Osaka University and Prof. Shiniya Kanemura for affiliation. 
TM was supposed to join  Osaka University as a postdoctoral fellow in April but 
delayed due to travel restrictions related to ongoing pandemic. 
TM also thanks National Taiwan University and Prof. Wei-Shu Hou for temporary visiting position with grant number MOST 106-2112-M-002-015-MY3.

\appendix
\section{EDMs}\label{app:edm}
For the EDM calculations, the following parametrization is also useful.
\begin{align}
\mathcal{L}_{\phi\bar{f}f} =-\phi\bar{f}\big(g^S_{\phi\bar{f}f}+i\gamma_5g^P_{\phi\bar{f}f}\big)f,
\end{align}
where $\phi=h,H,A$ and 
\begin{align}
g_{h\bar{f}f}^S &= \frac{1}{\sqrt{2}}\Big[\lambda_fs_\gamma+{\rm Re}\rho_{ff}c_\gamma\Big],\quad
g_{h\bar{f}f}^P = \frac{1}{\sqrt{2}}{\rm Im}\rho_{ff}c_\gamma, \\
g_{H\bar{f}f}^S &= \frac{1}{\sqrt{2}}\Big[\lambda_fc_\gamma-{\rm Re}\rho_{ff}s_\gamma\Big],\quad
g_{H\bar{f}f}^P = -\frac{1}{\sqrt{2}}{\rm Im}\rho_{ff}s_\gamma, \\
g_{A\bar{f}f}^S &= \pm \frac{1}{\sqrt{2}}{\rm Im}\rho_{ff},
\quad
g_{A\bar{f}f}^P = \mp\frac{1}{\sqrt{2}}{\rm Re}\rho_{ff}, \label{gAff}
\end{align}
where the upper sign is for up-type fermions and the lower for down-type fermions in Eq.~(\ref{gAff}).

Here we list the loop functions appearing in the EDM calculations in Sec.~\ref{sec:edms}.
\begin{align}
f(\tau) &= \frac{\tau}{2}\int_0^1dx~\frac{1-2x(1-x)}{x(1-x)-\tau}\ln\left(\frac{x(1-x)}{\tau}\right),\\
g(\tau) &= \frac{\tau}{2}\int_0^1dx~\frac{1}{x(1-x)-\tau}\ln\left(\frac{x(1-x)}{\tau}\right), \\
\mathcal{J}_W^V(m_\phi)
&=\frac{2m_W^2}{m_\phi^2-m_V^2}
\bigg[
	-\frac{1}{4}\left\{\left(6-\frac{m_V^2}{m_W^2}\right)+\left(1-\frac{m_V^2}{2m_W^2}\right)\frac{m_\phi^2}{m_W^2} \right\} \nonumber\\
&\hspace{4cm}\times\big(I_1(m_W,m_\phi)-I_1(m_W,m_V) \big) \nonumber\\
&\hspace{2.5cm}
	+\left\{\left(-4+\frac{m_V^2}{m_W^2}\right)+\frac{1}{4}\left(6-\frac{m_V^2}{m_W^2}\right)
	+\frac{1}{4}\left(1-\frac{m_V^2}{2m_W^2}\right)\frac{m_\phi^2}{m_W^2}\right\} \nonumber\\
&\hspace{4cm}\times\big(I_2(m_W,m_\phi)-I_2(m_W,m_V) \big)
\bigg], \\
J_1(\tau_{WH^\pm},\tau_{tH^\pm})
&=\int_0^1\frac{dx}{x}~(2-x)
\bigg[
Q_t(1-x)J\left(\tau_{WH^\pm}, \frac{\tau_{tH^\pm}}{x}\right)
+Q_bxJ\left(\tau_{WH^\pm}, \frac{\tau_{tH^\pm}}{x}\right)
\bigg],
\end{align}
where $\tau_{ij}=m_i^2/m_j^2$, $Q_t=2/3$, $Q_b=-1/3$ and
\begin{align}
I_1(m_1,m_2)  &= -2\frac{m_2^2}{m_1^2}f\left(\frac{m_1^2}{m_2^2}\right), \quad
I_2(m_1,m_2)  = -2\frac{m_2^2}{m_1^2}g\left(\frac{m_1^2}{m_2^2}\right), \\
J(a, b) &= \frac{1}{a-b}\left[\frac{a}{a-1}\ln a - \frac{b}{b-1}\ln b\right].
\end{align}

\end{document}